\documentclass[journal]{IEEEtran}
\usepackage{amsmath,amssymb,physics,tabularx,graphicx,multirow,hhline,algorithm,mathtools,algpseudocode,varwidth,color,soul,amsthm,cite,bbm,dsfont}
\usepackage[skins,breakable,hooks,theorems]{tcolorbox}

\algnewcommand{\algorithmicforeach}{\textbf{for each}}
\algdef{SE}[FOR]{ForEach}{EndForEach}[1]
  {\algorithmicforeach\ #1\ \algorithmicdo}
  {\algorithmicend\ \algorithmicforeach}
\newsavebox{\ieeealgbox}

\ifCLASSINFOpdf
\else
\fi

\tcbset{
  highlight math style={
    colback=yellow,
    arc=0pt,
    outer arc=0pt,
    boxrule=0pt,
    top=2pt,
    bottom=2pt,
    left=2pt,
    right=2pt,
  }
}

\ifCLASSOPTIONcompsoc \usepackage[caption=false,font=normalsize,labelfont=sf,textfont=sf]{subfig}
\else
\usepackage[caption=false,font=footnotesize]{subfig}
\fi

\begin{document}
\algnewcommand{\algorithmicgoto}{\textbf{go to}}%
\algnewcommand{\Goto}[1]{\algorithmicgoto~\ref{#1}}%
\title{Cyber-resilient Automatic Generation Control for Systems of AC Microgrids}

\author{Tong Huang,~\IEEEmembership{Member,~IEEE,}
		Dan Wu,~\IEEEmembership{Member,~IEEE,}
        and Marija Ili\'{c},~\IEEEmembership{Life Fellow,~IEEE}
        \thanks{
        This work was supported in part by the U.S. Department of Energy’s Office of Energy Efficiency and Renewable Energy (EERE) through the Solar Energy Technologies Office (SETO) under Grant DE-EE0009031.

Tong Huang is with the Department of Electrical and Computer Engineering, San Diego State University, San Diego, CA, 92182, USA; e-mail: thuang7@sdsu.edu. Dan Wu and Marija Ili\'{c} are with the Laborarory for Information and Decision Systems at Massachusetts Institute of Technology, Cambridge, MA 02139, USA; e-mail: \{danwumit, ilic\}@mit.edu.
}
        }



\maketitle

\begin{abstract}
In this paper we propose a  co-design of the secondary frequency regulation in systems of AC microgrids and its cyber security solutions. We term the secondary frequency regulator a Micro-Automatic Generation Control ($\mu$AGC) for highlighting its  same functionality as  the AGC  in bulk power systems.
We identify sensory challenges and cyber threats facing the $\mu$AGC. To address the sensory challenges, we introduce a new microgrid model by exploiting the rank-one deficiency property of microgrid dynamics. This model is used to pose an optimal $\mu$AGC control problem  that is easily implemented, because it does not require fast frequency measurements.
An end-to-end cyber security solution to the False Data Injection (FDI) attack detection and mitigation is developed for the proposed $\mu$AGC. The front-end barrier of applying off-the-shelf algorithms for cyber attack detection is removed by introducing a data-driven modeling approach. Finally, we propose an observer-based corrective control for an
islanded microgrid and a collaborative mitigation scheme in systems of AC microgrids. We demonstrate a collaborative role of systems of microgrids during cyber attacks. The performance of the proposed cyber-resilient $\mu$AGC is tested in a system of two networked microgrids.

\end{abstract}


\IEEEpeerreviewmaketitle{}

\section{Introduction}

The recent decade has witnessed several large-scale electricity outages due to low-probability but high-impact events, such as extreme weather. Examples include the 2012 Hurricane Sandy power outage, the 2020 California rotating electricity outage caused by extreme heat wave \cite{CISO2020}, and the 2021 Texas power crisis resulting from unusually severe winter storms \cite{Texas2021}. During these extreme weather-related events, the bulk transmission systems failed to supply sufficient electric energy to distribution systems, causing millions of businesses and homes to lose their electricity supply \cite{CISO2020,Texas2021}. One promising solution to enhance resilience of distribution systems to these extreme events is to integrate Distributed Energy Resources (DERs), e.g., solar panels, micro-turbines, and energy storage, to the distribution systems. However, a large-scale deployment of DERs in the distribution systems introduces unprecedented complexity and the need for extensive  Distribution Management Systems (DMSs) \cite{9559389,9951598}. To reduce the management complexity, a distribution system with massive DERs can be configured to be a system of microgrids. Neighboring loads, DERs and their grid infrastructure are clustered into an AC or DC, small-scale power system, i.e., a microgrid. One microgrid can either operate autonomously in an \emph{islanded mode} \cite{8750828}, or connect to its host distribution system in a \emph{grid-connected mode} \cite{8750828}, or as an interactive  network with its neighboring microgrids and form a system of microgrids that does not physically connect to their host distribution system in a \emph{hybrid mode}. Each microgrid is managed by the Microgrid Management System ($\mu$MS). In such a configuration, the management burden of the DMS is significantly reduced, as the DMS only needs to coordinate several $\mu$MS \cite{9559389,9951598,guerrero2010:hierarchical, 8750828}.

For an AC microgrid, frequency regulation is an important functional  block in the $\mu$MS design, expecially when the microgrid enters the islanded or hybrid mode. Similar to bulk transmission systems, frequencies at the microgrid can be regulated in a hierarchical manner. The bottom layer of the frequency regulation entails local controllers in DERs, e.g., power controllers of inverter-based resources (IBRs), and governors of small synchronous generators. These local controllers are termed the primary control of microgrids \cite{8750828,guerrero2010:hierarchical}. As the primary control can yield steady-state errors of frequencies from their nominal values after a disturbance, a secondary control layer is needed in microgrids to regain the nominal frequency \cite{simpson2015:secondary,choi2021:distributed}. Since the secondary frequency regulation in AC  microgrids shares a very similar objective with the Automatic Generation Control (AGC) of bulk transmission systems, we term the secondary frequency regulator in microgrids Micro-AGC ($\mu$AGC) to highlight its connection with the AGC that has served the bulk power systems for half of a century. This paper focuses on the $\mu$AGC design.

There is a large body of literature that addresses the $\mu$AGC design. These design schemes can be generally categorized into centralized, distributed, and decentralized control \cite{khayat2019:secondary}.  \emph{Centralized} $\mu$AGC relies on a monopolistic decision-making platform that gathers measurements from the DERs under control, computes regulation signals, and dispatches these signals to the DERs \cite{lopes2006:defining,tsikalakis2011:centralized}. \emph{Distributed} $\mu$AGC employs cooperative control policies among neighboring DERs to achieve certain objectives. It can be further classified into three sub-groups: averaging-based methods \cite{simpson2015:secondary}, distributed-consensus-based methods \cite{manaffam2017:intelligent}, and event-triggering-based methods \cite{shi2019:distributed}. 
In \emph{decentralized} $\mu$AGC design, each DER participating in the $\mu$AGC recovers its own nominal frequency to a nominal value independently. The washout filter is investigated in \cite{han2017:analysis} and shown to be equivalent to a decentralized secondary control. In \cite{khayat2018:decentralized}, a decentralized linear–quadratic regulator (LQR) control design is proposed to achieve frequency restoration in an optimal way. Another idea to achieve decentralized frequency regulation is based on state estimation. The estimation-based $\mu$AGC \cite{lou2017:decentralised} employs the system states estimated locally instead of the true measurements to generate control signals.

However, a system of microgrids equipped with the $\mu$AGC design mentioned above may be still vulnerable to physical disturbances and/or cyber anomalies. Compared with high-voltage transmission systems, a microgrid is more sensitive to physical disturbances, e.g., load/renewable power fluctuations, and IBR connection and re-connection, due to its small scale and the zero/low inertia of its generation units \cite{8750828,5456209}. These disturbances cause large, and fast fluctuations of microgrid frequencies. The frequency changes induced by the disturbances may not be captured in an accurate and timely manner by the frequency sensors (shall
be shown in Section \ref{subs:sensory_challenge}). However, almost all the above-mentioned $\mu$AGC design schemes utilize frequency measurements  to make decisions. It is assumed that the frequency information can be reported by DERs or it can be accurately measured in a fast manner. In practice, while the frequency information can be produced by the IBR power controller as a digital, internal control command \cite{4118327}, such a digital signal may not be available for a third-party $\mu$AGC\footnote{By the third-party $\mu$AGC, we mean that the $\mu$AGC service is not designed and implemented by the inverter manufacture.}, and it can only be indirectly obtained by external frequency sensors. As a result, the desirable control performance of the $\mu$AGC relying on accurate frequency measurements might not be attained, especially with fast fluctuations from renewable generation and load. Therefore, it is advantageous to develop a $\mu$AGC that does not depend on the frequency information when a microgrid enters the islanded or hybrid operating mode.

Another important consideration for the $\mu$AGC design is the cyber security. While the distributed and decentralized $\mu$AGC schemes reduce the cyber risks of the $\mu$AGC by sparsifying or eliminating long-distance communication links among DERs, these $\mu$AGC schemes may be still susceptible to the false data injection (FDI) attack where the sensors can be manipulated to report wrong information to the $\mu$AGC. Many references concern attenuating the influence of the FDI attacks \cite{tan2021:cyberattack}. One category of methods explores state estimation techniques \cite{hug2012:vulnerability}, for example, the Kalman filter method \cite{kurt2018:distributed} and optimization-based state estimation methods \cite{liu2014:detecting}. These methods either rely on the accuracy of measurements or heavy computations \cite{aluko2022:real}. To address these challenges, observer-based detection and control methods have been proposed \cite{tan2021:cyberattack,cecilia2021:addressing}. Recently, the unknown input observer (UIO) has found a broad range of applications in FDI attack detection and cyber-resilient control designs \cite{alhelou2019:deterministic,khalghani2020:resilient} since it can separate the external disturbances from the attack signals, yielding a more sensitive anomaly diagnosis. 
However, these cyber security solutions are built upon existing $\mu$AGC designs that require accurate frequency information. It is an open-ended question whether these solutions can extend to other control designs that require different information other than frequencies. While some cyber attack solutions \cite{7738534,76372} are designed for a general model that serves as an abstraction of many engineering systems, e.g., a state-space model, the model that should be used for the purpose of developing cyber solutions heavily depends on the information feeding its controller. For a specific controller, it might not be straightforward to derive a proper model that lends itself to the cyber solution design even for domain experts. Therefore, the development of $\mu$AGC and its cyber solution are interdependent.

To overcome these challenges in this paper we introduce a  co-design of $\mu$AGC and its cyber solutions. 
We identify the sensory challenges and cyber vulnerability facing the $\mu$AGC design.
To address the sensory challenges, we introduce a new microgrid modeling approach by exploiting the structure of microgrid dynamics, i.e., the property of rank-one deficiency. 
Such a modeling approach leads to an optimal $\mu$AGC that is easily implemented, as it does not require fast frequency measurements that can be hard to obtain \cite{rey2017:secondary}.
An end-to-end cyber security solution to FDI attack detection and mitigation is developed for the proposed $\mu$AGC. The front-end barrier of applying off-the-shelf algorithms for cyber attack detection is removed by introducing a data-driven modeling approach. Besides, we propose an observer-based corrective control for an
islanded microgrid and a collaborative mitigation scheme in systems of AC microgrids. Compared with the existing $\mu$AGC design, the contribution of this paper is summarized as follows: 1) The proposed $\mu$AGC can regulate microgrid frequencies in the presence of both load/renewable power fluctuations and FDI cyber attacks; 2) Compared with the $\mu$AGC design that requires internal digital signals of DERs or fast frequency measurements, the proposed $\mu$AGC is more practically implementable, as it only requires real-time real power which can be measured fast and accurately; 3) Compared with the method that is specifically designed for one type of DERs \cite{khayat2018:decentralized}, the proposed $\mu$AGC design can address systems of microgrids with heterogeneous DERs; and 4) We demonstrate the role of systems of microgrids in terms of attenuating the influence of cyber attacks.

The rest of this paper is organized as follows: Section \ref{sec:challenge} presents the challenges facing the $\mu$AGC design; Section \ref{sec:SectionIII} introduces the new microgrid modeling approach that is used for designing the $\mu$AGC in the presence of fast load/renewable power fluctuations; 
Section \ref{sec:cyber_attack} demonstrates the cyber security solutions to the proposed $\mu$AGC; and Section \ref{sec:case_study} tests the proposed $\mu$AGC and its cyber security solutions.
\section{New Modeling for $\mu$AGC Design}
\label{sec:challenge}
This section starts with describing the dynamics of systems of microgrids in the conventional state space where voltage phase angles and frequencies are used  as the state variables. Then we point out the limitations of conventional $\mu$AGC design.
We derive a new model which lends itself to more robust $\mu$AGC with respect to measurement inaccuracies and cyber attacks. 
\subsection{Conventional Microgrid Modeling for $\mu$AGC design}

\begin{figure}
    \centering
    \includegraphics[width = \linewidth]{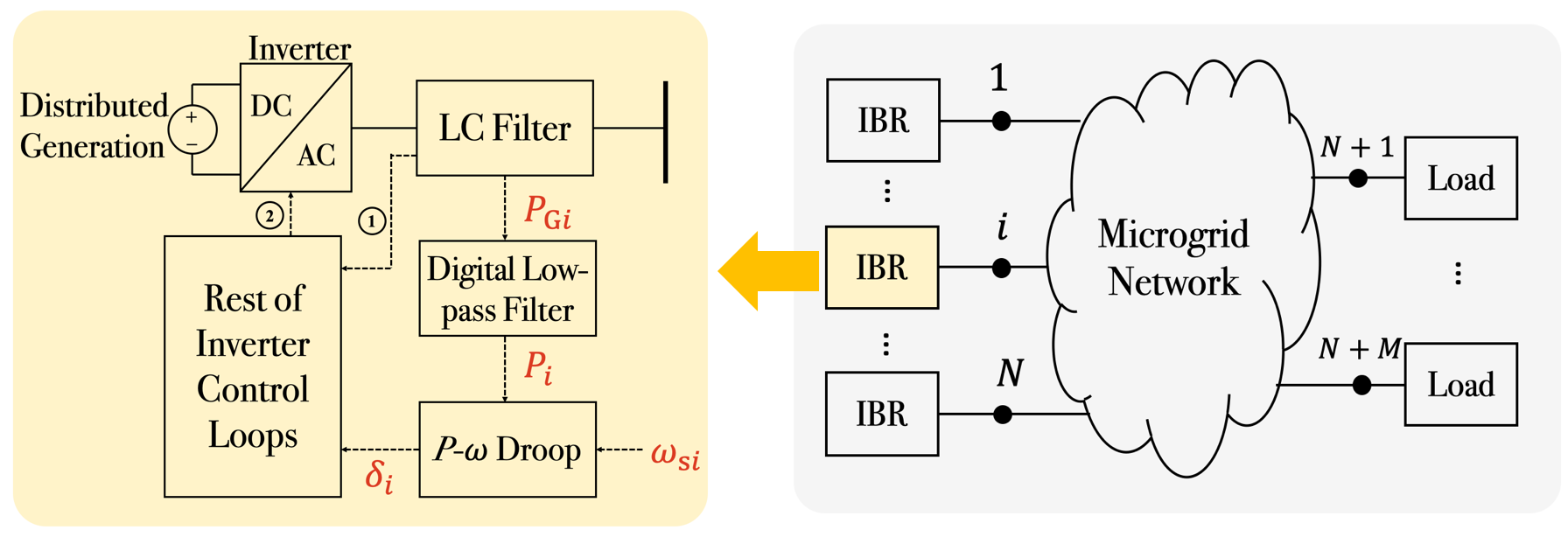}
    \caption{A microgrid with $N$ IBRs and $M$ loads and the control scheme of IBRs. The control variables of loops 1 and 2 are voltage and current measurements from the LC filter, and the modulation index, respectively.}
    \label{fig:IBR-CTR}
\end{figure}

We consider a microgrid with $N$ IBRs and $M$ loads. Without loss of generality, the IBRs are connected to the first $N$ nodes, and the loads and the point that networks with the neighboring microgrids locate at nodes $N+1, \ldots, N+M$, as shown in Figure \ref{fig:IBR-CTR}. The effect of the neighboring microgrids on the microgrid under study is modeled as load power injection.
\subsubsection{Nodal dynamics} Without $\mu$AGC and load/renewable power fluctuations, the primary control is designed such that the frequencies at all IBRs can be stabilized at $\omega_{\text{nom}}$ which is the desired/nominal frequency (i.e., 314 or 377 rad/s). The dynamics of the power calculator at the $i$-th IBR is
\begin{subequations}\label{eq:original_dynamic}
\begin{align} 
     &\dot{\delta}_i = \omega_i  - \omega_{\text{nom}} \label{eq:angle}\\
     &\dot{P_i} = - \omega_{\text{c}i} P_i + \omega_{\text{c}i}P_{\text{G}i}\label{eq:filter}\\
     & \omega_i = \omega_{\text{s}i} - m_{\text{P}i}P_i \label{eq:droop}
\end{align} 
\end{subequations}
where $\delta_i$ is the internal voltage phase angle; $\omega_i$ is frequency; $P_{\text{G}i}$ is the instantaneous real power component \cite{4118327}; $P_i$ is the real power filtered by the digital filter; $\omega_{\text{c}i}$ is the cut-off frequency of the digital filter; $\omega_{\text{s}i}$ is the setpoint; and $m_{\text{P}i}$ is the droop coefficient. Equation \eqref{eq:angle} is introduced by definition; and Equations \eqref{eq:filter} and \eqref{eq:droop} result from the digital filter dynamics and the droop characteristic \cite{4118327}. Variables $P_{\text{G}i}$, $P_i$, $\delta_i$, and $\omega_{\text{s}i}$ are annotated in Figure \ref{fig:IBR-CTR}.
Since $\omega_i$ is the variable that is regulated by the $\mu$AGC, we eliminate $P_i$ by noting 
\begin{equation} \label{eq:small_eq}
     \dot{\omega}_i = -m_{\text{P}i} \dot{P}_i; \quad
     P_i = (\omega_{\text{s}i}-  \omega_i)/m_{\text{P}i}.
\end{equation}
The underlying assumption of \eqref{eq:small_eq} is that the change rate of setpoint $\dot{\omega}_{\text{s}i}$ is much slower than $\dot{\omega}_i$, i.e., $\dot{\omega}_{\text{s}i}\approx 0$. With \eqref{eq:small_eq}, we obtain the following dynamics for the $i$-th IBR where $\omega_i$ becomes a state variable:
\begin{subequations} \label{eq:power_cal_dynamics}
\begin{align}
     &\dot{\delta}_i = \omega_i- \omega_{\text{nom}}\\
     &\dot{\omega}_i = - \omega_{\text{c}i}\omega_i + \omega_{\text{c}i}\omega_{\text{s}i} - m_{\text{P}i}\omega_{\text{c}i} P_{\text{G}i}.
\end{align} 
\end{subequations}
Denote by $\delta_i^*$, $\omega_i^*$, and $P_{\text{G}i}^*$ the steady-state values of $\delta_i$, $\omega_i$, and $P_{\text{G}i}$, respectively. Suppose that there are no load/renewable power fluctuations. If we need $\omega_i$ to converge to $\omega_{\text{nom}}$, $\omega_{\text{s}i}$ should be set to $\omega_{\text{s}i}^*$ where $\omega_{\text{s}i}^* = \omega_{\text{nom}}+m_{\text{P}i}P_{\text{G}i}^*$.
    
Define $\Delta\delta_i = \delta_i - \delta_i^*$, $\Delta\omega_i=\omega_i - \omega_i^*$, $\Delta P_{\text{G}i} = P_{\text{G}i}-P_{\text{G}i}^*$, and $\Delta\omega_{\text{s}i} = \omega_{\text{s}i} - \omega_{\text{s}i}^*$.
Use $\Delta \mathbf{x}_i=[\Delta\delta_i$, $\Delta\omega_i]^{\top}$ as a state vector. Then, Equation \eqref{eq:power_cal_dynamics} becomes
\begin{equation} \label{eq:linear_IBR}
    \Delta\dot{\mathbf{x}}_i = A_i\Delta\mathbf{x}_i + B_{1i}\Delta\omega_{\text{s}i} + B_{2i}\Delta P_{\text{G}i}
\end{equation}
where
\begin{equation*}
    A_i = \begin{bmatrix}
    0 & 1\\
    0 & -\omega_{\text{c}i}
    \end{bmatrix},
    \quad
    B_{1i} =
    \begin{bmatrix}
    0 \\
    \omega_{\text{c}i}
    \end{bmatrix},
    \quad
    B_{2i} =
    \begin{bmatrix}
    0 \\
    -m_{\text{P}i} \omega_{\text{c}i}
    \end{bmatrix}.
\end{equation*}
The dynamics for $N$ IBRs can be described by
\begin{equation} \label{eq:state_space}
    \Delta \dot{\mathbf{x}} = A_{\text{G}}\Delta\mathbf{x} + B_1 \Delta \boldsymbol{\omega}_{\text{s}} + B_2 \Delta \mathbf{P}_{\text{G}}
\end{equation}
where
\begin{subequations}
    \begin{align*}
        &\Delta \mathbf{x} = [\Delta \mathbf{x}_1^{\top}, \ldots, \Delta \mathbf{x}_N^{\top}]^{\top}; \Delta \boldsymbol{\omega}_{\text{s}} = [\Delta\omega_{\text{s}1}, \ldots \Delta\omega_{\text{s}N}]^{\top};\\
        & \Delta \mathbf{P}_{\text{G}} = [P_{\text{G}1}, \ldots, P_{\text{G}N}]^{\top}; A_{\text{G}} = \text{diag}(A_1, \ldots, A_N);\\
        &B_1 = \text{diag}(B_{11}, \ldots, B_{1N});B_2 = \text{diag}(B_{21}, \ldots, B_{2N}).
    \end{align*}
\end{subequations}

\subsubsection{Network Constraints}
For the microgrid under study, 
its neighboring microgrids that can be networked with can be considered power injection. For simplicity, real power and voltage magnitudes are assumed to be decoupled, i.e., the voltage magnitudes $V_i$ for $i\in\{1, \ldots, N+M \}$ is constant. The IBRs and loads are interconnected via microgrid network which introduces algebraic constraints for $i\in\{1, 2, \ldots, N+M\}$:
\begin{equation}\label{eq:network_eqs}
		P_{\text{I}_i} - V_i^{*2}g_{ii} - \sum_{k=1, k\ne i}^{N+M} V^*_iV^*_kY_{ik} \cos(\delta_{ik} - \theta_{ik})=0
\end{equation}
where $P_{\text{I}_i}$ is the net real injection to node $i$; $g_{ii}$ denotes the self-conductance of node $i$; the admittance of the branch from the $i$-th to the $k$-th node is $Y_{ik} \angle{\theta_{ik}}$; $V^*_i$ is the nominal voltage magnitude at node $i$; and $\delta_{ik}$ is the voltage phase angle difference between nodes $i$ and $k$, i.e., $\delta_{ik} = \delta_i - \delta_k$. Denote by $\{P_{\text{I}i}^*, Q_{\text{I}i}^*,  V_i^*, \delta_i^*| i=1,2,\ldots, N+M\}$ the operating condition that satisfies constraints \eqref{eq:network_eqs}. Define the following vectors
\begin{subequations}
\begin{align*}
    &\Delta \mathbf{P}_{\text{I}}:=[\Delta P_{\text{I}1}, \ldots, \Delta P_{\text{I}N+M}]^{\top} = [\Delta \mathbf{P}_{\text{G}}^{\top}, \Delta \mathbf{P}_{\text{L}}^{\top}]^{\top}\\
    &\Delta \boldsymbol{\delta}:=[\Delta \delta _{1}, \ldots, \Delta \delta_{N+M}]^{\top} =[\Delta \boldsymbol{\delta}_{\text{G}}^{\top}, \Delta \boldsymbol{\delta}_{\text{L}}^{\top}]^{\top}
\end{align*}
\end{subequations}
where $\Delta P_{\text{I}i}$ and $\Delta \delta_i$ are the deviations of $P_{\text{I}i}$ and $\delta_i$ from $P_{\text{I}i}^*$ and $\delta_i^*$, respectively; $\Delta \mathbf{P}_{\text{G}}=[\Delta P_{\text{G}1}, \ldots, \Delta P_{\text{G}N}]^{\top}$; $\Delta \mathbf{P}_{\text{L}}=[\Delta P_{\text{I}N+1}, \ldots, \Delta P_{\text{I}N+M}]^{\top}$; $\Delta \boldsymbol{\delta}_{\text{G}}=[\Delta \delta _{1}, \ldots, \Delta \delta_{N}]^{\top}$; $\Delta \boldsymbol{\delta}_{\text{L}}=[\Delta \delta _{N+1}, \ldots, \Delta \delta_{N+M}]^{\top}$; and the subscript of ``I'' of $\Delta P_{\text{I}i}$ stands for ``Injection''. If $\Delta P_{\text{I}i}$ and $\Delta \delta_i$ for all $i$ are small, the relationship between $\Delta \mathbf{P}_{\text{I}}$ and $\Delta \boldsymbol{\delta}_{\text{I}}$ can be described by
\begin{equation} \label{eq:linear}
    \Delta \mathbf{P}_{\text{I}} = H\Delta \boldsymbol{\delta}
\end{equation}
where the entry $h^{(i,k)}$ at the $i$-th row and the $k$-th column of $H\in\mathbb{R}^{(N+M)\times (N+M)}$ can be obtained via
\begin{subequations}
\begin{align*}
    &h^{(i,k)} = V_i^*V_k^*Y_{ik} \sin(\delta_{ik}^*-\theta_{ik})  \quad \forall k\ne i\\
    &h^{(i,i)} = -\sum_{p = 1, p\ne i}^{N+M} V_i^* V_p^* Y_{ip} \sin(\delta_{ip}^* - \theta_{ip}).
\end{align*}
\end{subequations}
Equation \eqref{eq:linear} is equivalent to
\begin{equation}
    \begin{bmatrix}
        \Delta \mathbf{P}_{\text{G}}\\
        \Delta \mathbf{P}_{\text{L}}
    \end{bmatrix} =
    \begin{bmatrix}
    H_{\text{GG}} & H_{\text{GL}}\\
    H_{\text{LG}} & H_{\text{LL}}
    \end{bmatrix}
    \begin{bmatrix}
        \Delta \boldsymbol{\delta}_{\text{G}}\\
        \Delta \boldsymbol{\delta}_{\text{L}}
    \end{bmatrix}
\end{equation}
where $H_{\text{GG}}\in \mathbb{R}^{N\times N}$, $H_{\text{GL}}\in \mathbb{R}^{N\times M}$, $H_{\text{LG}}\in \mathbb{R}^{M\times N}$, and $H_{\text{LL}}\in \mathbb{R}^{M\times M}$.
Based on the above equation, $\Delta \mathbf{P}_{\text{G}}$ can be expressed as a function of $\Delta \boldsymbol{\delta}_{\text{G}}$ and $\Delta \mathbf{P}_{\text{L}}$, i.e.,
\begin{equation} \label{eq:PG}
	\Delta \mathbf{P}_{\text{G}} = (H_{\text{GG}}- H_{\text{GL}}H_{\text{LL}}^{-1}H_{\text{LG}}) \Delta \boldsymbol{\delta}_{\text{G}} + H_{\text{GL}} H_{\text{LL}}^{-1} \Delta \mathbf{P}_{\text{L}}.
\end{equation}

\subsubsection{System Dynamics}
Plugging \eqref{eq:PG} into \eqref{eq:state_space}, we have
\begin{equation} \label{eq:ss_load_fluctuate}
    \Delta \dot{\mathbf{x}} = A\Delta\mathbf{x} + B_1\Delta \boldsymbol{\omega}_{\text{s}} + F \Delta \mathbf{P}_{\text{L}}
\end{equation}
where $A=A_{\text{G}} +B_2(H_{\text{GG}}-H_{\text{GL}}H_{\text{LL}}^{-1}H_{\text{LG}})E$; and $F=B_2H_{\text{GL}}H_{\text{LL}}^{-1}$. In the expression of matrix $A$, matrix $E= \text{diag}(E_1, \ldots, E_N)$ where $E_i = [1, 0]$ for $i = 1, \ldots,N$. It is worth noting that dynamics \eqref{eq:ss_load_fluctuate} directly establish the connection between the frequency change $\Delta \omega_i$ and the load/tie-line flow fluctuations $\Delta \mathbf{P}_{\text{L}}$.



\subsubsection{Two Common Assumptions for Conventional $\mu$AGC Design}
The objective of $\mu$AGC is to drive the frequency deviations $\{\Delta \omega_i|i=1,\ldots,N\}$ to zero by tuning the setpoints of IBRs $\boldsymbol{\omega}_{\text{s}i}$ in the presence of load/renewable power fluctuations $\Delta \mathbf{P}_{\text{L}}$.
With such an objective, a centralized $\mu$AGC typically tunes the setpoints $\boldsymbol{\omega}_{\text{s}}$ of the IBRs under control by measuring the frequency at a critical node. A decentralized/distributed $\mu$AGC observes its local frequency and other variables, e.g., local real power $P_{\text{G}i}$, to tune its local setpoint $\omega_{\text{s}i}$. Note that these $\mu$AGC design schemes are generally built upon at least one of the following two assumptions: 1) An IBR can report its true frequency $\omega_i$ to its $\mu$AGC, or $\omega_i$ can be measured accurately by external sensors; and 2) The measurements feeding the secondary controllers are authentic. However, it is possible that neither assumption holds in practice. This will be elaborated in the following two subsections.


\subsection{Sensory Challenges of $\mu$AGC Design} \label{subs:sensory_challenge}
\begin{figure}[h!]
				\centering
				\subfloat[]{\includegraphics[width=1.7in]{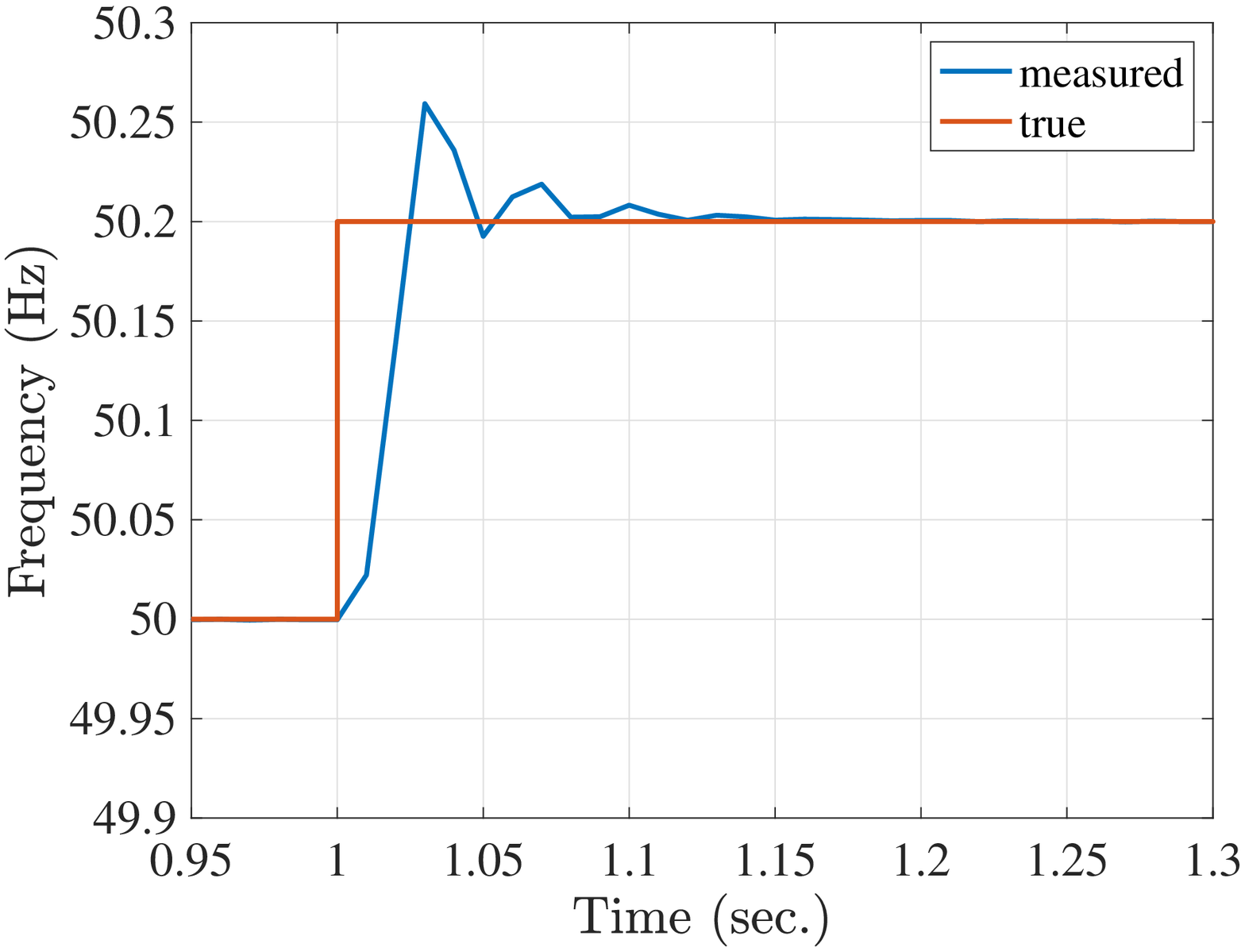}}
				\hfil
				\subfloat[]{\includegraphics[width=1.7in]{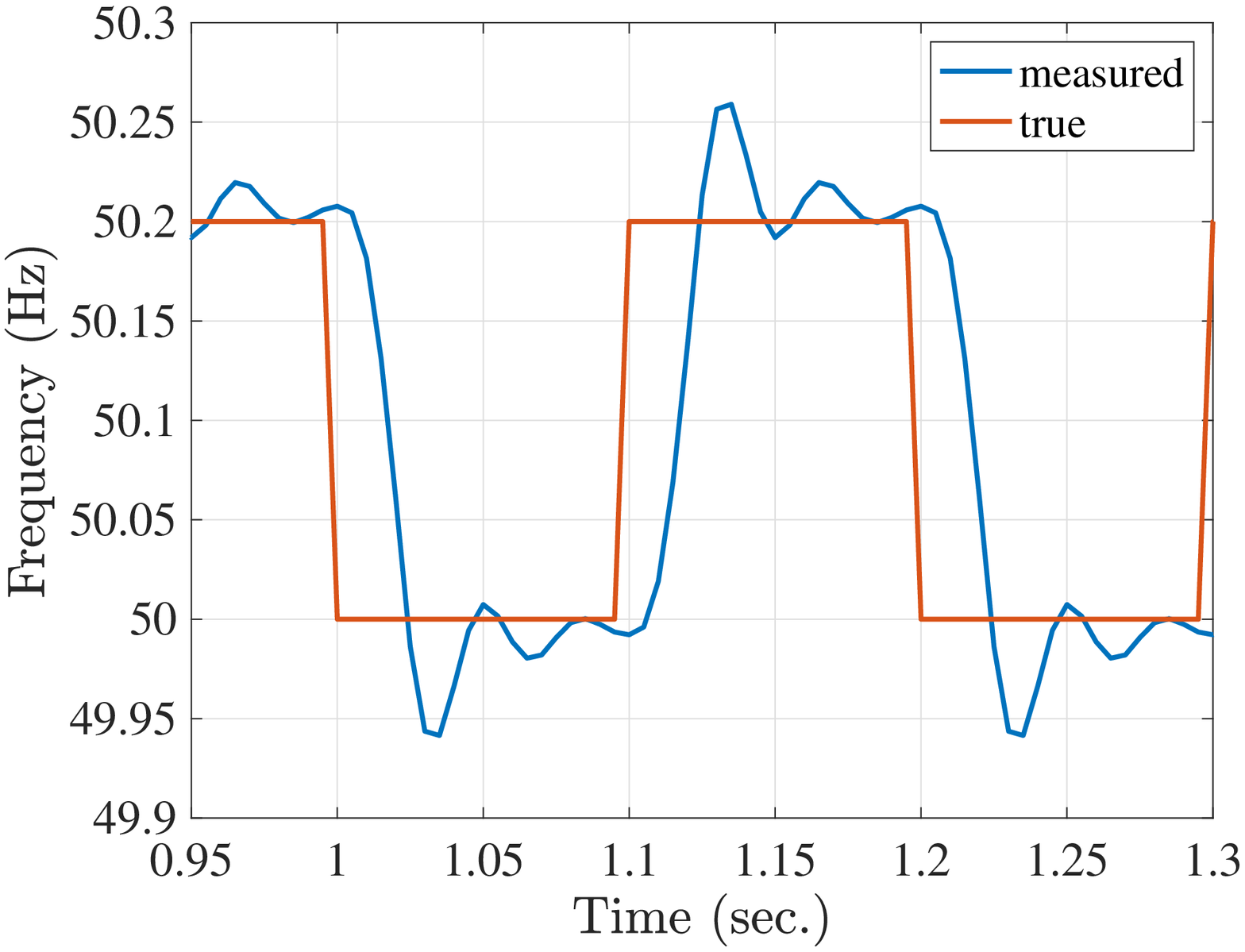}}
				\hfil
				\caption{(a) The true frequency can be measured if it changes slowly; (b) The true frequency cannot be accurately measured if it changes fast.}
				\label{fig: PMUFails}
			\end{figure}

The frequency $\omega_i$ is used as an input for many conventional $\mu$AGC to tune the setpoint $\boldsymbol{\omega}_{\text{s}i}$. For a commercial inverter, this variable is an internal, digital control command that might not be accessible for a third-party $\mu$AGC. One alternative solution in order to obtain $\omega_i$ is to measure the frequency of the fundamental component of terminal voltage at the $i$-th IBR. However, it is challenging to measure frequency both fast and accurately in practice. Figure \ref{fig: PMUFails} shows the sensory challenges when actual frequency changes fast and the frequency is required to be measured fast. In Figure \ref{fig: PMUFails}, the orange lines represent the true frequency of three-phase sinusoidal waves, while the blue curves shows the frequencies measured by a Simulink Phasor Measurement Unit (PMU) block \cite{PMUSimulink}. In Figure \ref{fig: PMUFails}(a), the true frequency changes at the $1$-st second, the measured frequency converges to the true frequency. The measurement error is acceptable if the PMU reports the frequency at a slow rate, say 10 samples per second ($10$ Hz). However, large measurement errors may exist from time $t=1$ s to $t=1.1$ s, if frequency is required to be reported every 0.01 seconds ($100$ Hz). Furthermore, as shown in Figure \ref{fig: PMUFails}(b), if the frequency keeps fluctuating, the large measurement errors can persist when a high sampling rate, say, $100$ Hz, is required.

One obvious question is why a high sampling rate of measurements is necessary for $\mu$AGC, compared with the conventional AGC in bulk transmission systems which issues control commands every $2$-$4$ seconds \cite{8345676,EPRI1234}. The reason is that the fast, large frequency fluctuations are more pronounced in small systems like microgrids than in the bulk transmission systems. Similar to the bulk transmission systems, there are volatile loads/renewables that keep perturbing the microgrid, which cause the microgrid frequencies to keep fluctuating. Compared with the transmission systems, the microgrid is more sensitive to these disturbances due to its smaller scale and the low inertia of its DERs \cite{8750828}. As a result, the microgrid frequencies may have large, fast fluctuations. To regulate the microgrid frequencies to their nominal values, the rate of issuing control commands should be much faster than the rate at which disturbances change. Otherwise, the frequencies may not be regulated. To \textcolor{black}{issue} the fast control commands, the fast and accurate measurements are needed. Since the frequency cannot be measured accurately in a fast manner, a conventional $\mu$AGC that takes measured frequency as inputs cannot achieve its desirable control performance. This motivates us to design a fast $\mu$AGC without using the fast measurements of frequencies. The solution  that addresses the sensory challenge is presented in Section \ref{sec:SectionIII}.

\subsection{Cyber Vulnerability of $\mu$AGC}

\begin{figure} [hbt]
        \centering
        \includegraphics[width = 3.5 in]{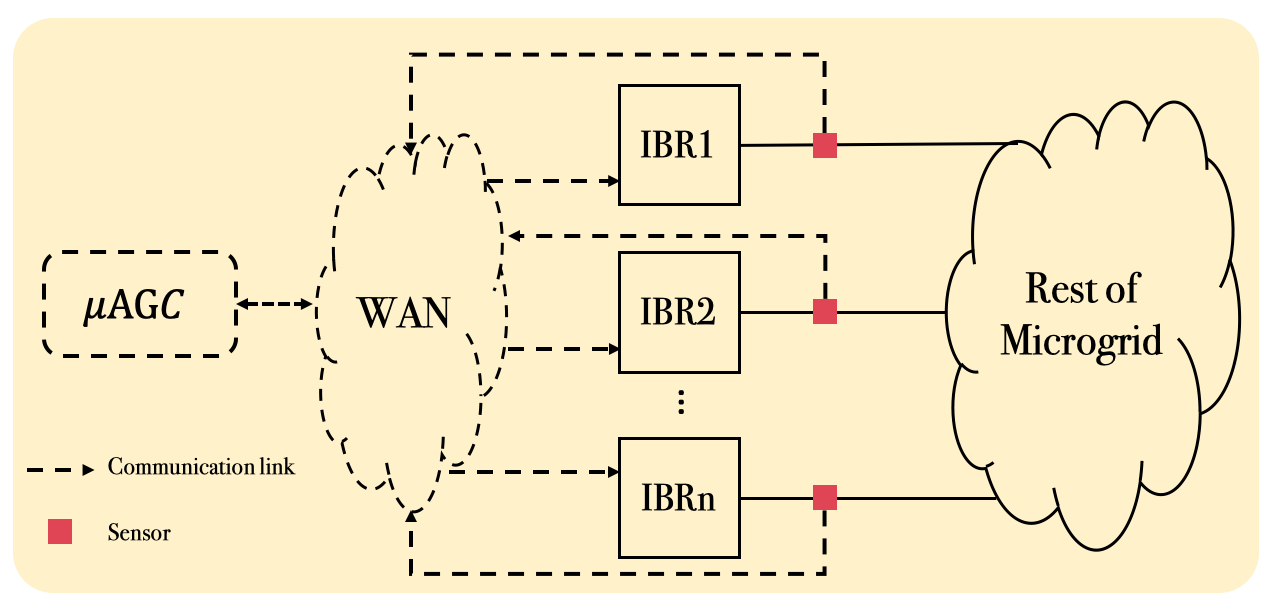}
        \caption{Communication architecture of a centralized $\mu$AGC}
        \label{fig:comm_arch}
    \end{figure}

A third-party $\mu$AGC coordinates IBRs based on sensor measurements. 
\textcolor{black}{Figure \ref{fig:comm_arch} shows the communication architecture of the proposed $\mu$AGC where the sensors that measure electrical variables send their measurements to $\mu$AGC via a Wide Area Network (WAN) \cite{ccliu_powergrid_review}. Various protocols can be applied to the WAN to establish wireless communication between the sensors and $\mu$AGC (See \cite{ccliu_powergrid_review} for more details of the protocols). FDI attacks can be launched by intruding into the WAN and/or manipulating the sensors. For most wireless communication, license-free ISM (industrial, scientific, and medical) radio band is used \cite{ccliu_powergrid_review}. As a result, the bandwidth can be legally accessed by attackers who aim to tamper with the information transmitted through the communication channel \cite{ccliu_powergrid_review}. The risks of such a type of attacks can be reduced by latest cryptographic mechanisms \cite{ccliu_powergrid_review}. This paper focuses on the FDI attacks launched via the sensors shown in Figure \ref{fig:comm_arch}. Such a type of attacks is feasible in real world. For example, Reference \cite{UCI_hall} has designed and demonstrated a device that can introduce FDI attack on Hall sensors in a non-intrusive manner by changing an external magnetic field.
 If such cyber attacks occur in the sensors shown in Figure \ref{fig:comm_arch},} all control commands from the $\mu$AGC can compromise the safety and efficiency of the microgrids. 
 For example, a cyber attacker can blind the $\mu$AGC by launching a \emph{replay attack}, i.e., the actual measurements $\mathbf{y}$ is replaced by a sequence of pre-recorded $\mathbf{y}$ \cite{8345676}. As another example, an adversary may compromise the efficiency of the $\mu$AGC by superposing random noise upon the actual measurements $\mathbf{y}$. Such a FDI attack is termed the \emph{noise injection} attack \cite{8345676}. \textcolor{black}{A large body of literature presents cyber attack models and objectives in the context of power grids (see \cite{liang2016:review, ccliu_AMI} and the references therein) and power electronics devices (See \cite{CPS_SmartInverters_survey} and the references therein).} A key question is how to design the $\mu$AGC that is resilient to the FDI attacks on the information feeding the $\mu$AGC. Section \ref{sec:cyber_attack} proposes a potential answer to this question.

\section{Fast Micro-Automatic Generation Control} \label{sec:SectionIII}


This section introduces a new model that exploits the rank-deficiency property of the microgrid dynamics and describes the microgrid behaviors in a new state space. Such a model lends itself to address the sensory challenge in the presence of fast load/renewable power fluctuations. Based on the model in the new space, we introduce a fast $\mu$AGC for a system of microgrids.

\subsection{New State Space Modeling of Microgrid Dynamics}
We describe the microgrid dynamics in a new state-space by exploiting the rank-deficiency property of the system matrix $A_i$ in \eqref{eq:linear_IBR}. Such a property is used for analyzing bulk transmission systems in our previous work \cite{ilic1993simple, liu2013large}. 
It is obvious that $A_i$ is rank deficient, i.e., one of eigenvalues of $A_i$ is zero. Denote by $T_i \in \mathbb{R}^{1\times 2}$ the left eigenvector associated with the zero eigenvalue of matrix $A_i$. By definition, we have $T_i A_i = \mathbf{0}$. One choice of $T_i$ is
\begin{equation} \label{eq:transformation}
    T_i = [\omega_{\text{c}i}, 1]. 
\end{equation}
Define a scalar $z_i$ as follows:
\begin{equation}
    z_i:= T_i\Delta \mathbf{x}_i.
\end{equation}
Multiplying both sides of \eqref{eq:linear_IBR} by $T_i$, we obtain
\begin{equation} \label{eq:z_dynamics}
    \dot{z}_i = \omega_{\text{c}i} \Delta \omega_{\text{s}i} - m_{\text{P}i} \omega_{\text{c}i}\Delta P_{\text{G}i}.
\end{equation}

Here, we show that the scalar $z_i$ lends itself to design a fast $\mu$AGC. The reason lies in the following two observations:
\begin{itemize}
    \item \textbf{Observation 1:} $z_i$ reflects the frequency regulation objective: a zero $\dot{z}_i$ implies that $\Delta \omega_i$ tends to zero.
    \item \textbf{Observation 2:} Computing $z_i$ only requires $P_{\text{G}i}$, suggesting that $z_i$ can be obtained in a fast manner.
\end{itemize}
\emph{Observation 1} results from Equations \eqref{eq:linear_IBR} and \eqref{eq:z_dynamics} that lead to
\begin{equation}
    \Delta \dot{\omega}_i = - \omega_{\text{c}}\Delta\omega_i + \dot{z}_i. 
\end{equation}
With a zero $\dot{z}_i$, $\Delta \omega_i(t)$ tends to zero, as time $t$ tends to infinity, given any intial condition $\Delta \omega_i(0)$. This is because the scalar system $\Delta \dot{\omega}_i = - \omega_{\text{c}}\Delta\omega_i$ is asymptotically stable, where $\omega_{\text{c}}>0$. Observation 1 essentially connects the frequency regulation with the scalar $z_i$, and it suggests that the frequency at the $i$-th IBR can be regulated, if the corresponding $\dot{z}_i$ equals $0$. \emph{Observation 2} is obtained by integrating both sides of \eqref{eq:z_dynamics}:
\begin{equation} \label{eq:z_i_int}
    z_i(t) = \omega_{\text{c}i}\int_0^{t}\left(\Delta \omega_{\text{s}i}(\tau)-m_{\text{P}i}\Delta P_{\text{G}i}(\tau)\right) d\tau + z_i(0)
\end{equation}
where $\Delta \omega_{\text{s}i}$ is the control command produced by the secondary controller; $\Delta P_{\text{G}i}$ can be measured; and the initial condition $z_i(0)$ can be set to zero for engineering implementation. 

Essentially, \eqref{eq:z_i_int} suggests that if the $\mu$AGC makes decisions by observing $z_i$ in the presence of fast load/renewable power fluctuations, measuring fast $\Delta P_{\text{G}i}$ suffices, and it is unnecessary to have fast frequency measurements. A natural question is: Compared with frequency, why can the instantaneous power $\Delta P_{\text{G}i}$ be measured both fast and accurately? In practice, frequency and instantaneous power are computed based on instantaneous voltage and current measurements. The instantaneous voltage (current) can be measured by a voltage (current) transducer. After being filtered by a low-pass filter, the analog signal of the voltage (current) can be digitized by an analog-to-digital circuit \cite{7265090}. With the digital voltage/current signal, the frequency of the voltage/current can be estimated by various algorithms \cite{7086070}, e.g., phase-locked loops \cite{5575433}, discrete Fourier transform \cite{7086070}, and least-squares techniques \cite{216849}. Before these signal processing techniques accurately track the true frequency, there exists a transient process where the frequency estimated does not match the true frequency. The waveform from 1 sec. to 1.15 sec. in Figure \ref{fig: PMUFails}(a) of the manuscript shows such a transient process. As shown in Figure \ref{fig: PMUFails}(b), the concern is that when the true frequency changes during the sensor transient process, these signal processing techniques cannot accurately estimate the frequency anymore. However, with the digital voltage and current signals with a high sampling rate (e.g., $30.7$ kHz \cite{relay_manual}), the instantaneous power can be computed almost instantaneously using an algebraic equation\footnote{The algebraic equation  is the definition of the instantaneous three-phase power $p_{3\phi}$: $p_{3\phi}(t) = v_a(t)i_a(t)+v_b(t)i_b(t)+v_c(t)i_c(t)$ where $\{v_a, v_b, v_c\}$ and $\{i_a, i_b, i_c\}$ are the instantaneous, three-phase voltages and currents, respectively.}. The operation of computing the instantaneous power will not incur the transient process that appears when the frequencies are computed. Therefore, the instantaneous power can be measured accurately, even when the true power fluctuates fast. The sensory bottleneck described in Section \ref{subs:sensory_challenge} is overcome by introducing a new variable $z_i$.

It is worth noting that the modeling approach is applicable to other types of DERs that possesses the rank-one deficiency property. This is because this rank deficiency is a direct consequence of conservation of energy \cite{ilic1993simple, liu2013large}. Our earlier work \cite{8974033} addresses synchronous machines using the same modeling approach. Such a modeling approach lends itself to a  $\mu$AGC  design for microgrids with heterogeneous energy resources. The $\mu$AGC design for such a kind of microgrids cannot be formulated by some control methods that are specifically designed for IBRs, such as \cite{khayat2018:decentralized}.
\subsection{Optimal Frequency Regulation} \label{subs:OFR}
This subsection introduces an optimal $\mu$AGC that tunes the IBR setpoints $\Delta \omega_i$ by observing the scalars $z_i$. To drive the frequency deviations to zero, based on Observation 1, one possible control law is
\begin{equation} \label{eq: ctrl_law1}
    \Delta \omega_{\text{s}i} = m_{\text{P}i}\Delta P_{\text{G}i} \quad \forall i=1, \ldots, N. 
\end{equation}
According to \eqref{eq:z_dynamics}, the control law \eqref{eq: ctrl_law1} leads to a zero $\dot{z_i}$. It follows that $\Delta \omega_{\text{s}i}$ tends to zero, based on Observation 1.
It is worth noting that the control law \eqref{eq: ctrl_law1} can be implemented in a decentralized manner, as the setpoint for the $i$-th IBR is computed only by its local measurement $P_{\text{G}i}$ for $i=1,2, \ldots,N$. However, the decentralized control law \eqref{eq: ctrl_law1} might not achieve high economical efficiency, since different IBRs incurs different generation costs. For example, we may expect the cheaper IBRs to generate more real power in order to regulate the microgrid frequencies. Therefore, coordination among IBRs is needed in order to regulate frequencies at the minimal costs. Such coordination is achieved by formulating the frequency regulation problem into the following optimal control problem in the $z_i$ space:
\begin{equation} \label{eq:opt_z}
    \begin{aligned}
    \min_{\Delta \boldsymbol{\omega}_{\text{s}}}\quad \frac{1}{2} \int_0^{\infty} &\left[\mathbf{z}(t)^{\top}Q\mathbf{z}(t)+\Delta\boldsymbol{\omega}_{\text{s}}(t)^{\top}R\Delta\boldsymbol{\omega}_{\text{s}}(t)\right]dt\\
    \text{s.t.} \quad &\Delta \dot{\mathbf{x}}(t) = A\Delta\mathbf{x}(t) + B_1\Delta \boldsymbol{\omega}_{\text{s}}(t)
    \end{aligned}
\end{equation}
where $\mathbf{z}=[z_1, \ldots, z_N]^{\top}$; matrices $Q \in \mathbb{R}^{N\times N}$ and $R \in \mathbb{R}^{N\times N}$ are positive definite. In the frequency regulation problem in this paper, both matrices $Q$ and $R$ are diagonal. Denote by $q_i$ and $r_i$ the $i$-th diagonal entry of $Q$ and $R$, respectively. A relatively large $q_i$ suggests relatively higher quality of service is required by the end users at node $i$. A relatively large $r_i$ suggests a relatively higher generation cost at the $i$-th IBR. Based on transformation \eqref{eq:transformation}, $\mathbf{z}$ can be expressed in terms of states $\Delta\mathbf{x}$ in the conventional space
\begin{equation} \label{eq:z_vector}
    \mathbf{z} = T \Delta\mathbf{x}
\end{equation}
where $T = \text{diag}(T_1, \ldots, T_N)$. Plugging \eqref{eq:z_vector} into the objective function of the optimization \eqref{eq:opt_z} gives the standard formulation of the LQR problem:
\begin{equation} \label{eq:opt_x}
    \begin{aligned}
    \min_{\Delta \boldsymbol{\omega}_{\text{s}}}\quad \frac{1}{2} \int_0^{\infty} &\left[\Delta\mathbf{x}(t)^{\top}Q'\Delta\mathbf{x}(t)+\Delta\boldsymbol{\omega}_{\text{s}}(t)^{\top}R\Delta\boldsymbol{\omega}_{\text{s}}(t)\right]dt\\
    \text{s.t.} \quad &\Delta \dot{\mathbf{x}}(t) = A\Delta\mathbf{x}(t) + B_1\Delta \boldsymbol{\omega}_{\text{s}}(t)
    \end{aligned}
\end{equation}
where $Q'=T^{\top}QT$. The solution to \eqref{eq:opt_x} has the following form: $\Delta \boldsymbol{\omega}_{\text{s}} = -K'\Delta \mathbf{x}$,
where $K'$ can be obtained a process that has been standardized in some software toolboxes, e.g., the function $\texttt{lqr}$ in MATLAB. As discussed earlier, it is more desirable to feed $\mu$AGC the transformed states $\mathbf{z}$, rather than $\Delta\mathbf{x}$ which includes the frequency deviation $\Delta \omega_i$ that is hard to measure at a fast rate. Therefore, the control law is
\begin{equation} \label{eq:Kz}
    \Delta \boldsymbol{\omega}_{\text{s}} = -K\mathbf{z}.
\end{equation}
where matrix $K$ is chosen such that $K\mathbf{z} \approx K'\Delta \mathbf{x}$, and one possible choice of $K$ is
$K=K'T^{\top}(TT^{\top})^{-1}$ \cite{liu2013large}.
Note that the control law \eqref{eq:Kz} tunes $\Delta \boldsymbol{\omega}_{\text{s}}$ by observing $\mathbf{z}$ or by measuring $P_{\text{G}i}$ for $i=1, 2, \ldots,N$. Such a control law is immune to the FDI attacks on frequency or phase angle measurements. However, it is vulnerable to cyber attacks on $P_{\text{G}i}$. The cyber vulnerability of the control law \eqref{eq:Kz} is addressed in the following section.
\section{Cyber Anomaly Detection and Correction}
\label{sec:cyber_attack}
To compute $\mathbf{z}$, the real power $P_{\text{G}i}$ for all IBRs is needed to be measured and reported to a $\mu$AGC. It is possible that a malicious adversary can compromise the microgrid by launching the FDI attacks (described in Section II-C) on the $P_{\text{G}i}$. If this happens, the control command $\Delta \boldsymbol{\omega}_{\text{s}}$ issued by the $\mu$AGC may fail to stabilize or to regulate the frequency, compromising the microgrid's safety and the quality of electricity service. This section introduces an end-to-end scheme that can enhance the microgrid resilience to cyber attack on the critical measurements. The key idea of the scheme is to leverage a computational model that represents  the dynamics of the microgrid, for the purpose of predicting $P_{\text{G}i}$. The main challenge is how to obtain the computational model that bridges the healthy information $\Delta \boldsymbol{\omega}_{\text{s}}$ to $P_{\text{G}i}$ that might be under cyber attacks. We first present the computational model that serves as the prerequisite of applying existing cyber attack detection algorithms. Next, we propose a strategy that can mitigate the impact of cyber attacks.
\subsection{Computational Model for Cyber Attack Mitigation}

Without load/renewable power fluctuations, the real power $\Delta\mathbf{P}_{\text{G}}$ used for computing $\mathbf{z}$ in \eqref{eq:Kz} can be predicted by $\Delta \boldsymbol{\omega}_{\text{s}}$ according to the following discrete state space model with the sample time $\Delta t$:
\begin{subequations}\label{eq:PG_predict_part}
    \begin{align}
        &\hat{\mathbf{x}}[k] = A'\hat{\mathbf{x}}[k-1] + B' \Delta \boldsymbol{\omega}_{\text{s}}[k-1]\label{eq:PG_predict_part1}\\
        &\Delta\hat{\mathbf{P}}_{\text{G}}[k] = C'\hat{\mathbf{x}}[k] 
    \end{align}
\end{subequations}
where $\Delta\hat{\mathbf{P}}_{\text{G}}$ is the predicted version of $\Delta\mathbf{P}_{\text{G}}[k]$;  $\hat{\mathbf{x}}[k]$ is an intermediate variable which can be obtained by using \eqref{eq:PG_predict_part1} recursively if $\hat{\mathbf{x}}[0]$ is given; and $A'$, $B'$, and $C'$ are the system, input, and output matrices.

One challenge of using \eqref{eq:PG_predict_part} to predict $\Delta\mathbf{P}_{\text{G}}$ is how to obtain the matrices $A'$, $B'$, and $C'$. We leverage a system identification technique (i.e., $\texttt{N4SID}$ \cite{van1993n4sid}) to learn these matrices using the experimental measurements of $\Delta \boldsymbol{\omega}_{\text{s}}$ and $\Delta\mathbf{P}_{\text{G}}$. Implementing the $\texttt{N4SID}$ technique requires: 1) input-output data; and 2) the order of the system \eqref{eq:PG_predict_part}. 
\subsubsection{Input-output data} 
Define a rectangular function $\texttt{rect}(t)$ as follows
\begin{equation}
    \texttt{rect}(t) = \mathds{1}_{t\ge 0} - \mathds{1}_{t-\Delta t'\ge 0}
\end{equation}
where $\Delta t'>\Delta t$ denotes the width of the rectangular pulse; and $\mathds{1}_{t\ge 0}$ is an indicator function, i.e., 
\begin{equation*}
    \mathds{1}_{t\ge 0}=
    \begin{cases}
    1 & t \ge 0\\
    0 & t < 0.
    \end{cases}
\end{equation*}
The input signal $\Delta \boldsymbol{\omega}_{\text{s}}'[k]$ for system identification can be sampled from the following continuous function with sampling time $\Delta t$
\begin{equation}
    l(t) = \sum_{k=0}^{\lceil t/\Delta t'\rceil} \alpha_k \texttt{rect}(t-k\Delta t')
\end{equation}
where $\lceil\cdot\rceil$ is the ceiling function; the random scalar $\alpha_k$ is uniformly distributed in $[-\beta, \beta]$ whence $\beta$ is a small positive scalar. By feeding the microgrid the designed setpoint sequence $\mathcal{S}= \{\Delta \boldsymbol{\omega}_{\text{s}}'[k]|k= 0, 1, \ldots, K_0\}$, the response of $\Delta \mathbf{P}_{\text{G}}$ can be measured at sample time $\Delta t$. The input sequence $\mathcal{S}$ associated with the $\Delta \mathbf{P}_{\text{G}}$'s response is used  for implementing the $\texttt{N4SID}$ technique.

\subsubsection{Order Selection}
The optimal system order $d^*$ is obtained by trial and error. Suppose that there a set $\mathcal{D}$ that collects some candidate system orders, say, $\mathcal{D}:=\{1,2, \ldots, 10\}$. For $d\in \mathcal{D}$, the $\texttt{N4SID}$ technique can return the identified $A'$, $B'$, and $C'$. With these matrices and the input data $\mathcal{O}$, $\Delta \mathbf{P}_{\text{G}}$'s prediction $\Delta \hat{\mathbf{P}}_{\text{G}d}$ can be computed. The performance of the identified system with the order $d$ can be quantified by
\begin{equation}
    \eta_d = \frac{1}{K_0+1}\sum_{k=0}^{K_0}\norm{\Delta \hat{\mathbf{P}}_{\text{G}d}-\Delta \mathbf{P}_{\text{G}}}_2.
\end{equation}
The optimal order $d^*$ is chosen such that the smallest $\eta_d$ is attained, viz., $\eta^*=\text{arg}\min_{d\in\mathcal{D}}\eta_d$.

\subsection{Cyber Attack Detection} \label{subs: cyber_attack_detection}
The state-space model \eqref{eq:PG_predict_part} can be used in some cyber attack detection algorithms, e.g., the dynamic watermarking approach \cite{7738534}, and the UIO method \cite{alhelou2019:deterministic}. As an example, we show how to use the dynamic watermarking approach to detect cyber attacks on $P_{\text{G}i}$. 

\begin{figure}
    \centering
    \includegraphics[width = 0.6\linewidth]{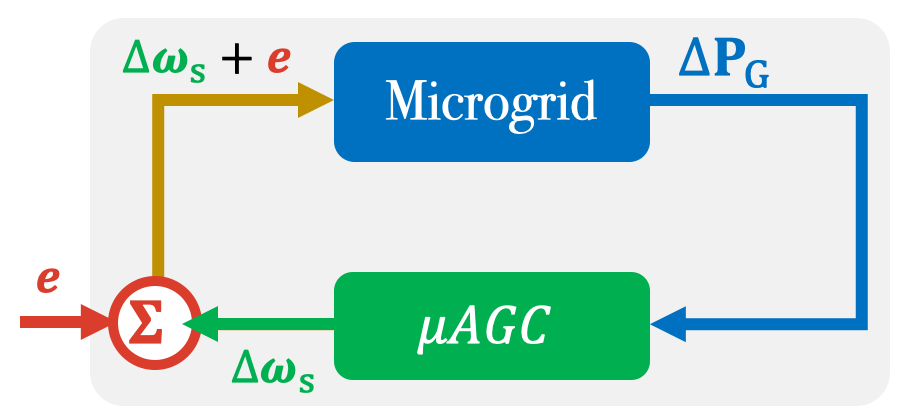}
    \caption{Basic idea of the dynamic watermarking approach}
    \label{fig:dynamic_watermark}
\end{figure}

Figure \ref{fig:dynamic_watermark} shows the basic idea of the dynamic watermarking approach. A small, secret watermark signal $\mathbf{e}[k] \in \mathbb{R}^{N}$ is superposed upon the IBRs' setpoints $\Delta\boldsymbol{\omega}_{\text{s}}$ at each time step $k$. The watermark signal possesses some known statistics, e.g. $\mathbf{e}[k] \sim \mathcal{N}(0, \Sigma)$ where matrix $\Sigma$ is the covariance matrix of $\mathbf{e}[k]$. The watermark $\mathbf{e}[k]$ impacts the measurement of $\Delta \mathbf{P}_{\text{G}}$. As a result, an authentic measurement of $\Delta \mathbf{P}_{\text{G}}$ should reflect the statistical properties concerning the watermark $\mathbf{e}[k]$. By subjecting the measurements of $\Delta \mathbf{P}_{\text{G}}$ to certain statistical tests, a wide range of cyber attacks on the measurement of $\Delta \mathbf{P}_{\text{G}}$ can be detected \cite{7738534,8345676}. 


The detailed mechanism of the dynamic watermarking approach is elaborated as follows.
Denote by $\Delta\mathbf{P}_{\text{G}}'$ the measurements of the real power $\Delta\mathbf{P}_{\text{G}}$ that are received by the $\mu$AGC. If there are no cyber attacks on $\Delta\mathbf{P}_{\text{G}}$, $\Delta\mathbf{P}_{\text{G}}' = \Delta\mathbf{P}_{\text{G}}$,  with the true sequence of $\{\Delta \boldsymbol{\omega}_{\text{s}}[i]\}$ and $\{\mathbf{e}[i]\}$ for $i=0,1,\ldots, k-1$, one can approximately predict $\Delta\mathbf{P}_{\text{G}}[k]$ based on
\begin{subequations}\label{eq:PG_predict_whole}
    \begin{align}
        &\hat{\mathbf{x}}[k] = A'\hat{\mathbf{x}}[k-1] + B' (\Delta \boldsymbol{\omega}_{\text{s}}[k-1] + \boldsymbol{e}[k-1])\label{eq:PG_predict1}\\
        &\Delta\hat{\mathbf{P}}_{\text{G}}[k] = C'\hat{\mathbf{x}}[k] \label{eq:PG_predict2}.
    \end{align}
\end{subequations}
Define
\begin{equation} \label{eq:innovation}
    \boldsymbol{\nu}[k]:=\Delta\mathbf{P}_{\text{G}}'[k]-\Delta\hat{\mathbf{P}}_{\text{G}}[k].
\end{equation}
For a microgrid without cyber attack, i.e., $\Delta\mathbf{P}_{\text{G}}'[k] = \Delta\mathbf{P}_{\text{G}}[k]$, we can obtain a long sequence $\{\Delta\mathbf{P}'_{\text{G}}[k]|k=1, \ldots,W\}$ where $W$ is a large integer. The expectation $\boldsymbol{\mu}_{\nu}^*$ and covariance matrix $\Sigma_{\nu}^*$ of $\boldsymbol{\nu}$ in the microgrid without cyber attack can be approximated by $\hat{\boldsymbol{\mu}}_{\nu}^*$ and $\hat{\Sigma}_{\nu}^*$ where
\begin{subequations}
    \begin{align}
        &\hat{\boldsymbol{\mu}}_{\nu}^* = \frac{1}{W} \sum_{k=1}^W \boldsymbol{\nu}[k]\\
        &\hat{\Sigma}_{\nu}^* =\frac{1}{W}\sum_{k=1}^W (\boldsymbol{\nu}[k]-\hat{\boldsymbol{\mu}}_{\nu}^*)(\boldsymbol{\nu}[k]-\hat{\boldsymbol{\mu}}_{\nu}^*)^{\top}.
    \end{align}
\end{subequations}

For the microgrid that might be under cyber attacks where $\Delta\mathbf{P}_{\text{G}}'[k]$ may not be equal to $\Delta\mathbf{P}_{\text{G}}[k]$, denote by $\hat{\boldsymbol{\mu}}_{\nu}$ and $\hat{\Sigma}_{\nu}$ the estimated expectation and covariance matrix of $\boldsymbol{\nu}$, respectively. Let $\mathcal{M}_{\rho}$ and $\hat{\mathcal{M}}_{\rho}$ collect $\Delta\mathbf{P}'_{\text{G}}$ and its prediction $\Delta\hat{\mathbf{P}}_{\text{G}}$ from time $t=(\rho-W+1)\Delta t$ to $t=\rho\Delta t$, viz.,
\begin{subequations}
    \begin{align}
        &\mathcal{M}_{\rho}:= \{\Delta\mathbf{P}'_{\text{G}}[k]|k=\rho-W+1, \ldots,\rho\}\\
        &\hat{\mathcal{M}}_{\rho}:= \{\Delta\hat{\mathbf{P}}_{\text{G}}[k]|k=\rho-W+1, \ldots,\rho\}.
    \end{align}
\end{subequations}
$\hat{\boldsymbol{\mu}}_{\nu}$ and $\hat{\Sigma}_{\nu}$ can be estimated from $\mathcal{M}_{\rho}$ and $\hat{\mathcal{M}}_{\rho}$ via
\begin{subequations} \label{eq:estimation}
    \begin{align}
        &\hat{\boldsymbol{\mu}}_{\nu} = \frac{1}{W}\sum_{k=\rho-W+1}^{\rho} \boldsymbol{\nu}[k] \label{eq:mu_hat}\\
        &\hat{\Sigma}_{\nu} =\frac{1}{W}\sum_{k=\rho-W+1}^{\rho} (\boldsymbol{\nu}[k]-\hat{\boldsymbol{\mu}}_{\nu})(\boldsymbol{\nu}[k]-\hat{\boldsymbol{\mu}}_{\nu})^{\top}. \label{eq:sigma_hat}
    \end{align}
\end{subequations}
If $\Delta\mathbf{P}_{\text{G}}'[k] = \Delta\mathbf{P}_{\text{G}}[k]$, $\hat{\boldsymbol{\mu}}_{\nu}$ and $\hat{\Sigma}_{\nu}$ should be close to $\hat{\boldsymbol{\mu}}_{\nu}^*$ and $\hat{\Sigma}_{\nu}^*$, respectively, that is,
\begin{subequations} \label{eq:two_conditions}
    \begin{align}
        &\xi_1 =\norm{\hat{\boldsymbol{\mu}}_{\nu}-\hat{\boldsymbol{\mu}}_{\nu}^*}_2 < \epsilon_1 \label{eq:cdt1}
        \\
        &\xi_2 = |\text{tr}(\hat{\Sigma}_{\nu}-\hat{\Sigma}_{\nu}^*)|<\epsilon_2
        \label{eq:cdt2}
    \end{align}
\end{subequations}
where $\norm{\cdot}_2$ and $\text{tr}(\cdot)$ are the $\mathcal{L}_2$ norm and the trace operator, respectively; and $\epsilon_1$ and $\epsilon_2$ are positive numbers. Violation of either \eqref{eq:cdt1} or \eqref{eq:cdt2} or both suggests cyber attacks, i.e., $\Delta\mathbf{P}_{\text{G}}[k]\ne \Delta\mathbf{P}'_{\text{G}}[k]$. The two conditions in \eqref{eq:two_conditions} can be checked in a moving-window fashion, which is summarized in Algorithm \ref{alg:1}. Function \texttt{DW} in Algorithm \ref{alg:1} returns a binary result: ``$\texttt{flag}=0$'' suggests that a cyber attack is not detected, while ``$\texttt{flag}=1$'' suggests that cyber attack is detected.


\begin{algorithm}
    \caption{Watermarking-based Cyber Attack Detection} \label{alg:1}
        \begin{algorithmic}[1]
            \Function{\tt DW}{$\mathcal{M}_{\rho-1}$,$\hat{\mathcal{M}}_{\rho-1}$, $\Delta \mathbf{P}_{\text{G}}'[\rho]$, $\Delta \boldsymbol{\omega}_{\text{s}}[\rho]$, $\mathbf{e}[\rho]$, $\hat{\mathbf{x}}'[\rho-1]$, $A'$, $B'$, $C'$, $\hat{\boldsymbol{\mu}}_{\nu}^*$, $\hat{\Sigma}_{\nu}^*$, $\epsilon_1$, $\epsilon_2$}
                \State Compute $\Delta\hat{\mathbf{P}}_{\text{G}}[\rho]$, $\hat{\mathbf{x}}$ by \eqref{eq:PG_predict_whole}
                \State $\hat{\mathcal{M}}_{\rho} \leftarrow \left\{\Delta \hat{\mathbf{P}}_{\text{G}}[\rho]\right\} \cup \hat{\mathcal{M}}_{\rho-1} -\left\{\Delta \hat{\mathbf{P}}_{\text{G}}[\rho-W]\right\}$
                \State $\mathcal{M}_{\rho} \leftarrow \left\{\Delta \mathbf{P}_{\text{G}}'[\rho]\right\} \cup \mathcal{M}_{\rho-1} -\left\{\Delta \mathbf{P}_{\text{G}}'[\rho-W]\right\}$
                \State Compute $\hat{\boldsymbol{\mu}}_{\nu}$, $\hat{\Sigma}_{\nu}$ using $\mathcal{M}_{\rho}$, $\hat{\mathcal{M}}_{\rho}$ based on \eqref{eq:innovation}, \eqref{eq:estimation}
                \State Obtain $\xi_1$ and $\xi_2$ according to \eqref{eq:two_conditions}
                \If{$(\xi_1<\epsilon_1)\land (\xi_2<\epsilon_2)$}
                \State $\texttt{flag}\leftarrow 0$ \algorithmiccomment{No attack is detected}
                \Else \State $\texttt{flag}\leftarrow 1$ \algorithmiccomment{Attack is detected}
                \EndIf
                \State \Return $\texttt{flag}$, $\mathcal{M}_{\rho}$, $\hat{\mathcal{M}}_{\rho}$, $\hat{\mathbf{x}}'[\rho]$ 
            \EndFunction
        \end{algorithmic}
\end{algorithm}

\subsection{Corrective Control}
This subsection addresses how to regulate frequency under cyber attacks on the real power measurements $\Delta \mathbf{P}_{\text{G}}$ feeding the proposed control law \eqref{eq:Kz}. Depending on whether there are microgrids that can be networked with the microgrid under cyber attacks, the $\mu$AGC under cyber attack can select one of the following two schemes to regulate frequencies.
\subsubsection{Observer-based Correction Scheme} \label{ssubsec: observer}
Once the cyber attack is detected and there is no neighboring microgrid that can be  networked with the microgrid under attack, the microgrid attacked can use the observer-based correction scheme to regulate its frequencies. The basic idea of such a scheme is to leverage an observer to predict the real power measurement $\Delta\mathbf{P_{\text{G}}}$ and to use the prediction $\Delta\hat{\mathbf{P}}_{\text{G}}$ to tune the setpoints $\Delta \boldsymbol{\omega}_{\text{s}}$ of IBRs.
Specifically, the observer is defined in \eqref{eq:PG_predict_part} \cite{WuACC2022}.
The control law in the observer-based correction scheme is
\begin{equation} \label{eq:Kz_hat}
    \Delta \boldsymbol{\omega}_{\text{s}} = -K
    \begin{bmatrix}
        \hat{z}_1, \hat{z}_2, \ldots, \hat{z}_N 
    \end{bmatrix}^{\top}.
\end{equation}
where $\hat{z}_i(t)=\omega_{\text{c}i}\int_0^{t}\left(\Delta \omega_{\text{s}i}(\tau)-m_{\text{P}i}\Delta \hat{P}_{\text{G}i}(\tau)\right) d\tau$.


\subsubsection{Collaborative Correction Scheme} \label{subsubs:networked}
The control law \eqref{eq: ctrl_law1} is blind to load/renewable power fluctuations. As a result, the frequencies cannot be regulated at the nominal frequency if $\Delta \mathbf{P}_{\text{L}}\ne 0$. This issue can be addressed by the collaborative correction scheme presented here. Suppose that there are two islanded microgrids. Each microgrid has its own $\mu$AGC designed based on the control law \eqref{eq:Kz}. Assume that $\mu$AGC$_1$ ($\mu$AGC$_2$) is for Microgrid 1 (Microgrid 2). Suppose that microgrid $1$ is under cyber attack on its $\Delta P_{\text{G}i}$ that is used for computing $z_i$. Once the cyber attack is detected at Microgrid $1$ by some cyber attack detection algorithms, e.g., the dynamic watermarking approach, $\mu$AGC$_1$ will be disabled, and Microgrid $1$ can be networked with Microgrid $2$ by closing the tie line  between Microgrids $1$ and $2$. As will be shown in Section \ref{sec:case_study}, $\mu$AGC$_2$ can regulate the frequencies at all nodes of both Microgrids $1$ and $2$, while  $\mu$AGC$_2$ makes decisions only based on the local measurements at Microgrid $2$. 

The overall procedure described throughout Sections \ref{sec:SectionIII} and \ref{sec:cyber_attack} is summarized in Algorithm \ref{alg:2} where set $\mathcal{I}_{\rho-1}$ collects the input for the function $ \texttt{DW}$ in Algorithm \ref{alg:1}, i.e.,
\begin{equation}
    \begin{aligned}
        \mathcal{I}_{\rho-1}:=&\{ \mathcal{M}_{\rho-1},\hat{\mathcal{M}}_{\rho-1},\Delta \mathbf{P}_{\text{G}}'[\rho],\Delta \boldsymbol{\omega}_{\text{s}}[\rho],\mathbf{e}[\rho],\\
        &\hat{\mathbf{x}}'[\rho-1],A',B',C',\hat{\boldsymbol{\mu}}_{\nu}^*,\hat{\Sigma}_{\nu}^*,\epsilon_1,\epsilon_2\}.
    \end{aligned}
\end{equation}

\begin{algorithm}
    \caption{Cyber Resilient $\mu$AGC} \label{alg:2}
        \begin{algorithmic}[1]
        \State $\mathcal{M}_W \leftarrow \cup_{k=1}^W \{\Delta \mathbf{P}_{\text{G}}'[k]\}$
        \While{$\rho = W+1, W+2, \ldots$}
            \State $\texttt{flag}, \mathcal{M}_{\rho}, \hat{\mathcal{M}}_{\rho}, \mathbf{x}'[\rho]\leftarrow\texttt{DW}(\mathcal{I}_{\rho-1})$
            \If{$\texttt{flag}=0$}
            \State Compute $\Delta \boldsymbol{\omega}_{\text{s}}[\rho]$ via \eqref{eq:z_i_int} and \eqref{eq:Kz}
            \Else
            \If{Neighboring microgrids are unavailable}
            \State Compute $\Delta \boldsymbol{\omega}_{\text{s}}[\rho]$ via \eqref{eq:PG_predict_part} and \eqref{eq:Kz_hat}. \State $\mathbf{e}[\rho]\leftarrow0$ \algorithmiccomment{Stop injecting watermark}
            \Else
            \State $\Delta \boldsymbol{\omega}_{\text{s}}[\rho]\leftarrow 0$; $\mathbf{e}[\rho]\leftarrow0$
            \State Network with the neighboring microgrids
            \EndIf
            \EndIf
        \State\Return $\Delta \boldsymbol{\omega}_{\text{s}}[\rho]$
        \EndWhile
        \end{algorithmic}
\end{algorithm}
\section{Case Study}
\label{sec:case_study}
In this section, the proposed fast $\mu$AGC and the associated cyber anomaly detection and correction scheme are tested in a microgrid with three IBRs via Simulink simulations. We first provide a brief description for the software-based microgrid testbed. Then we show the performance of the fast controller that exploits the rank deficiency property of IBRs, under fast renewable/load power fluctuations. Finally, we present the performance of the cyber attack detection and correction scheme that is built upon the model developed in Section \ref{sec:cyber_attack}.
\subsection{Description of Software-based Microgrid Testbed}
\begin{figure}
    \centering
    \includegraphics[width = 0.7\linewidth]{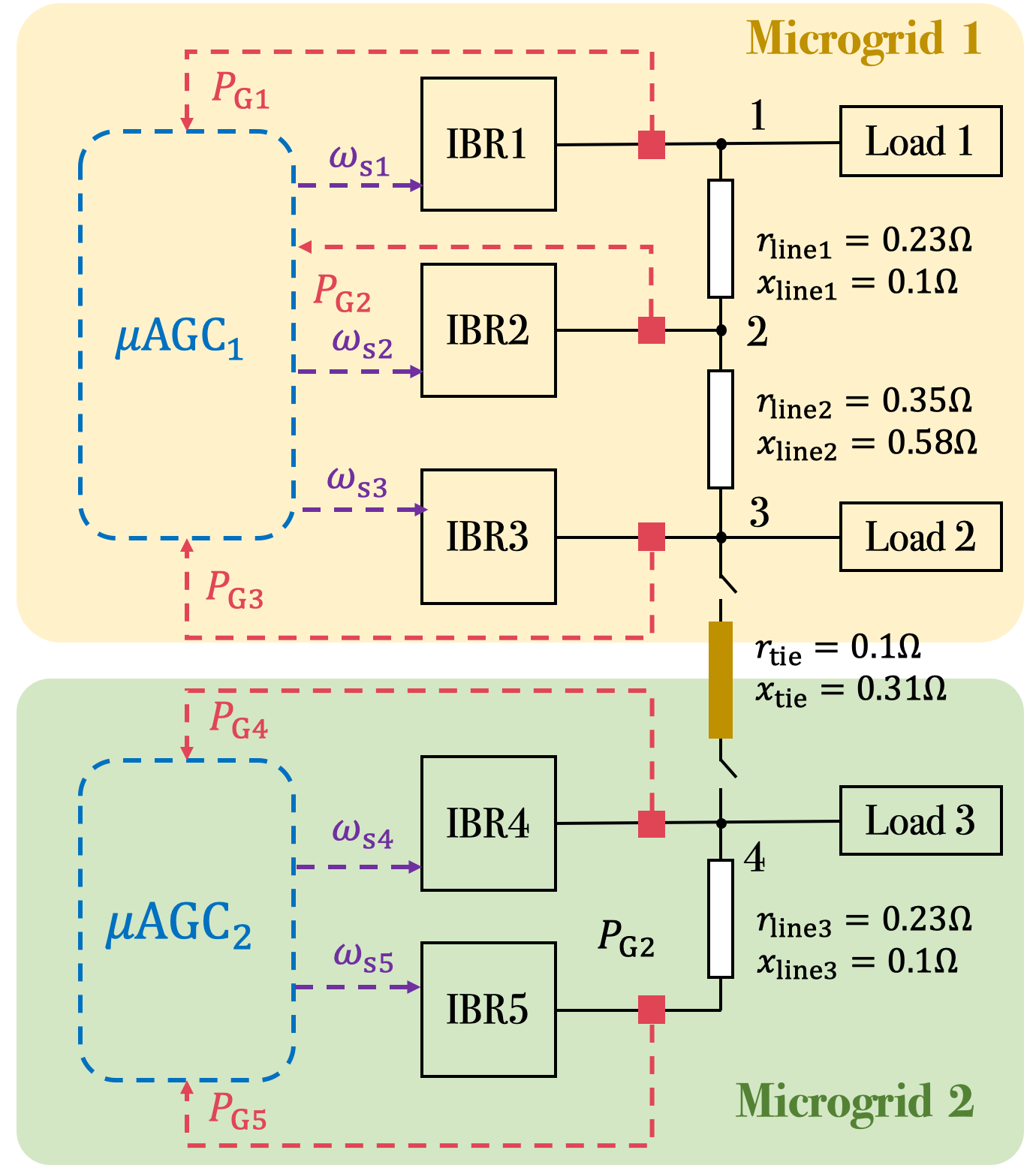}
    \caption{Physical and cyber architecture of two networked microgrids}
    \label{fig:two_networked_MG}
\end{figure}

Figure \ref{fig:two_networked_MG} illustrates the physical and cyber architecture of two networked microgrids that are used to test the proposed algorithms. For each IBR, its power calculator, voltage controller, current controller, and output filter are modeled according to \cite{4118327}. All IBRs have  the same parameters. Loads $1$, $2$, and $3$ are modeled by resistors with the values of $25$ $\Omega$, $20$ $\Omega$, and $33$ $\Omega$ for each phase. There is a tie line  that can network Microgrid $1$ with Microgrid $2$. The line parameters are annotated in Figure \ref{fig:two_networked_MG}. Each microgrid has its own $\mu$AGC which tunes the setpoint of each IBR in the local microgrid. 

\subsection{Control Performance without Cyber Attack}
\label{subs:nocyberattack}
Here the performance of the controller \eqref{eq:Kz} is tested. The secondary controller  $\mu$AGC$_1$ takes real power measurements $P_{\text{G}i}$ every $5$ ms and issues control commands $\omega_{\text{s}i}$ at the same rate, for $i=1,2,3$. The tie line in Figure \ref{fig:two_networked_MG} is open.
\begin{figure}[b]
				\centering
				\subfloat[]{\includegraphics[width=0.45\linewidth]{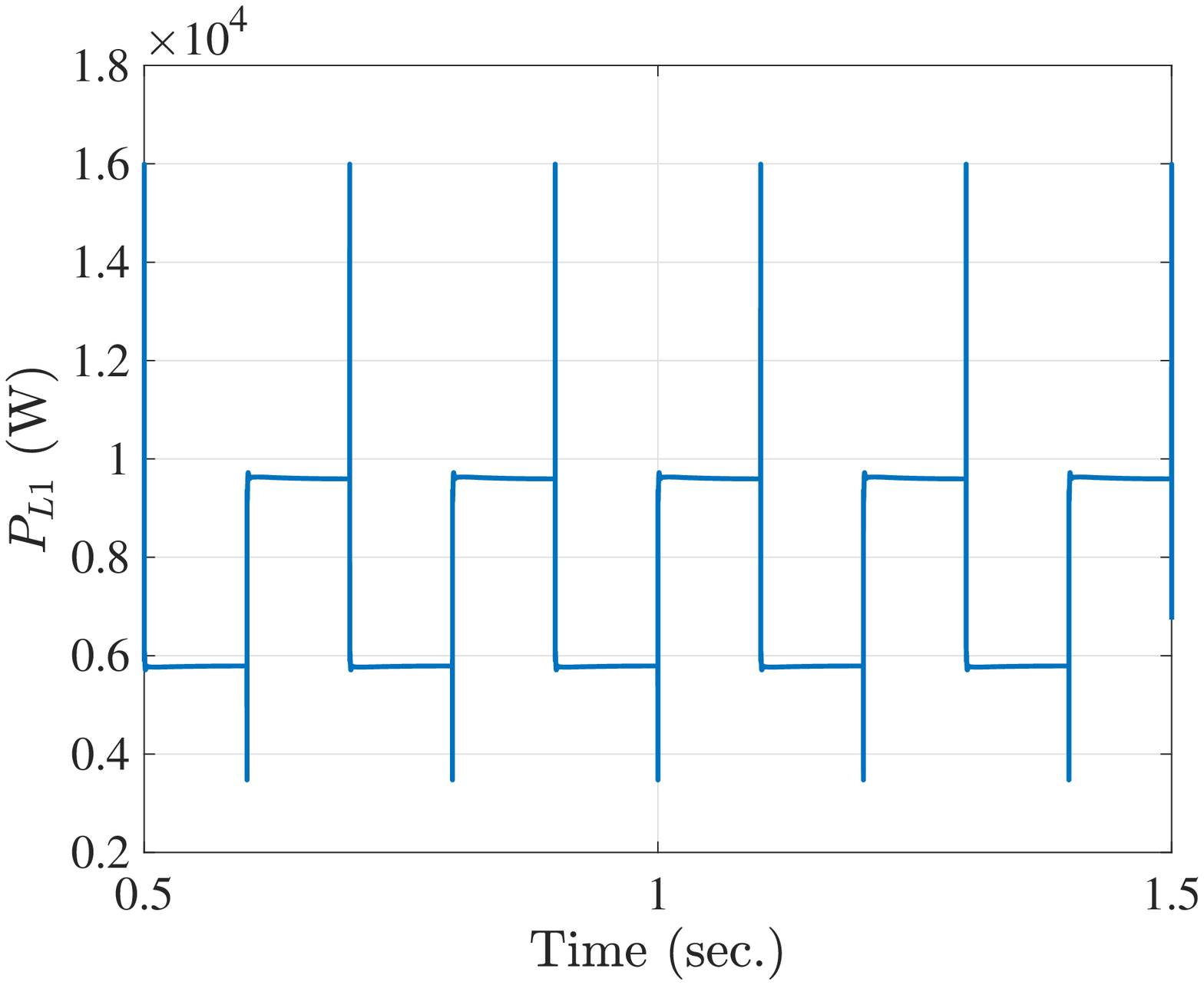}}
				\hfil
				\subfloat[]{\includegraphics[width=0.45\linewidth]{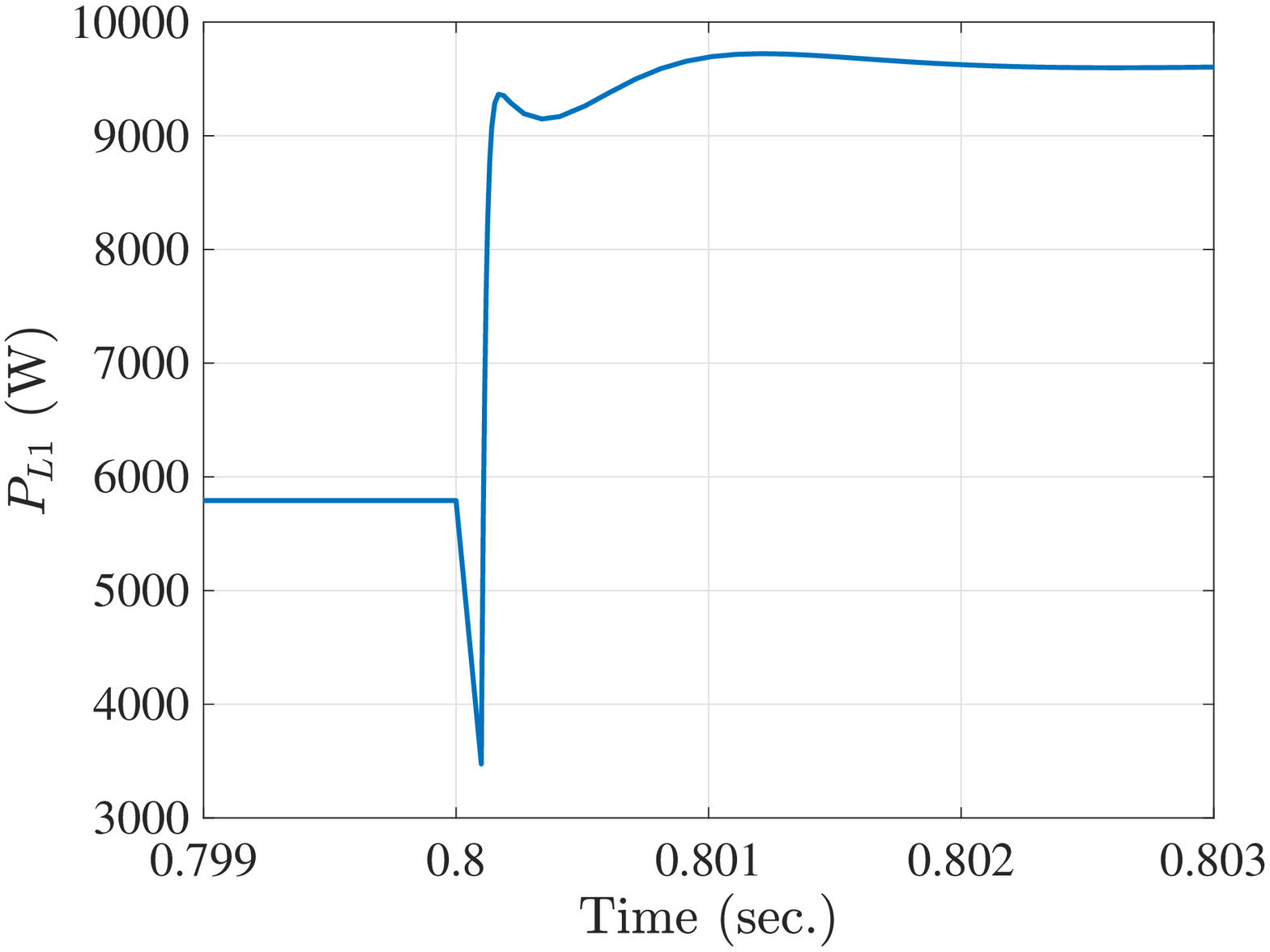}}
				\hfil
				\caption{(a) Real power measurement $P_{\text{L}1}$ at load 1; and (b) zoomed-in version of $P_{\text{L}1}$ from time $t = 0.799$ sec. to $t=0.803$ sec.}
				\label{fig: LoadFluctuate}
			\end{figure}

	\begin{figure*}[t]
				\centering
				\subfloat[]{\includegraphics[width=0.25\linewidth]{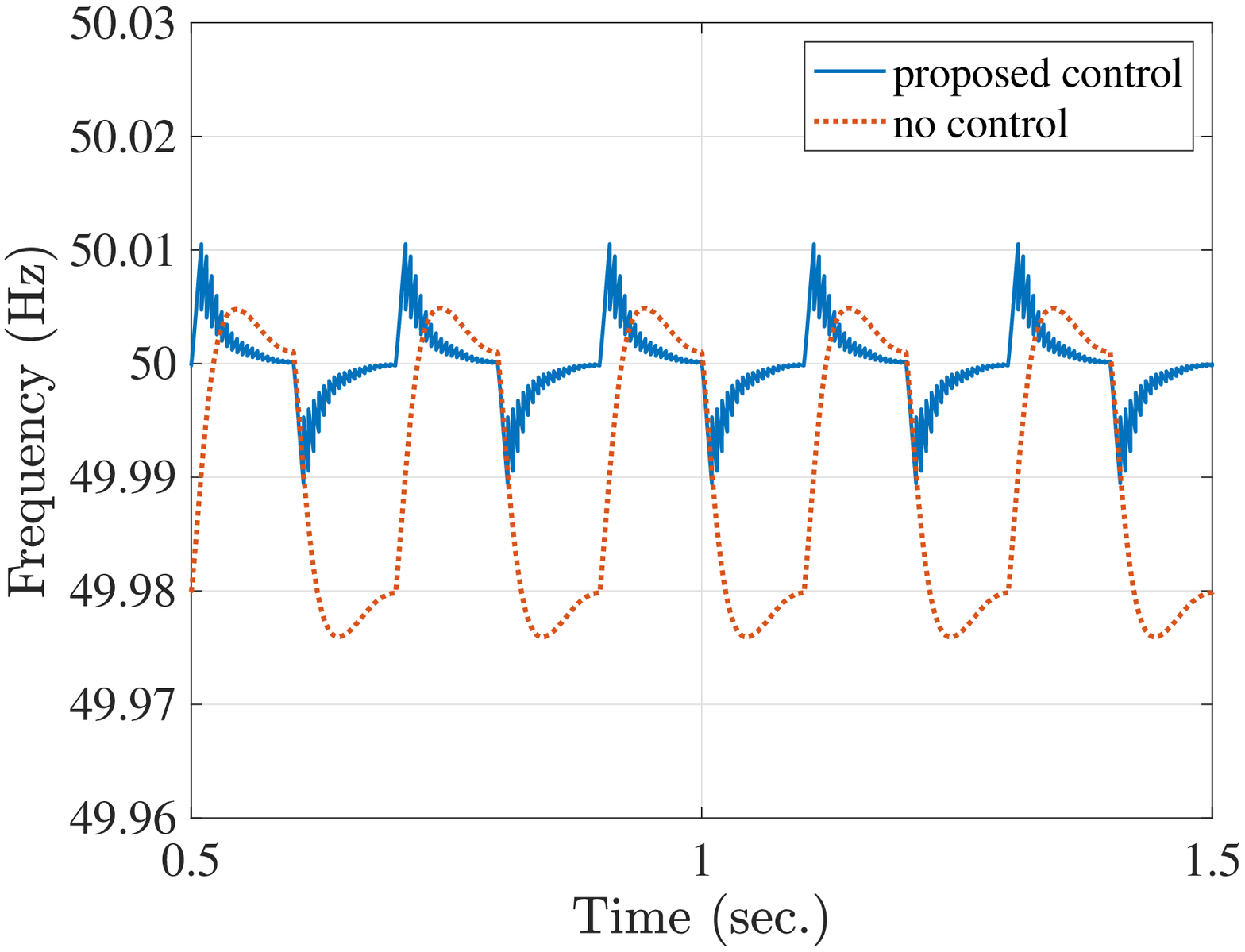}}
				\hfil
				\subfloat[]{\includegraphics[width=0.25\linewidth]{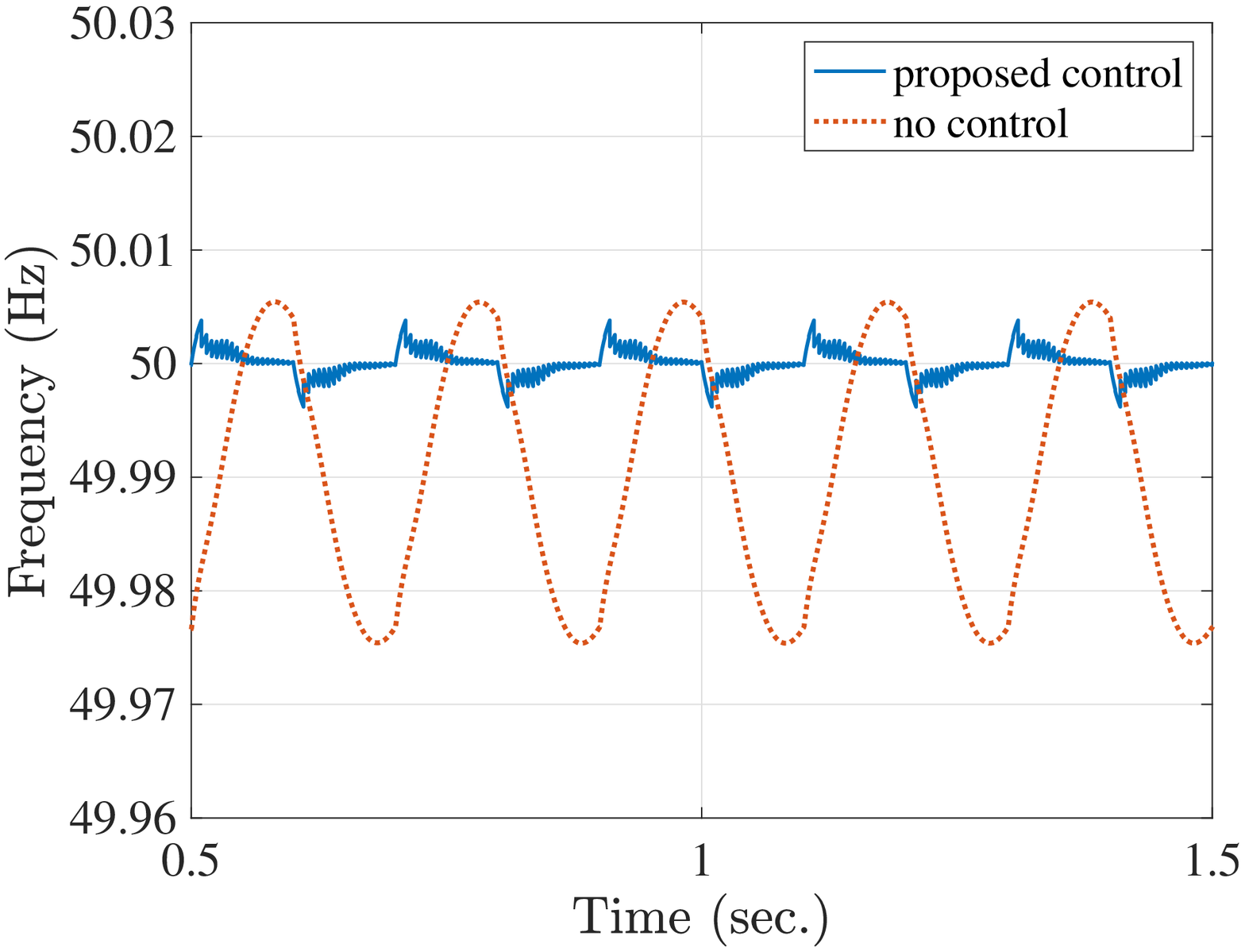}}
				\hfil
				\subfloat[]{\includegraphics[width=0.25\linewidth]{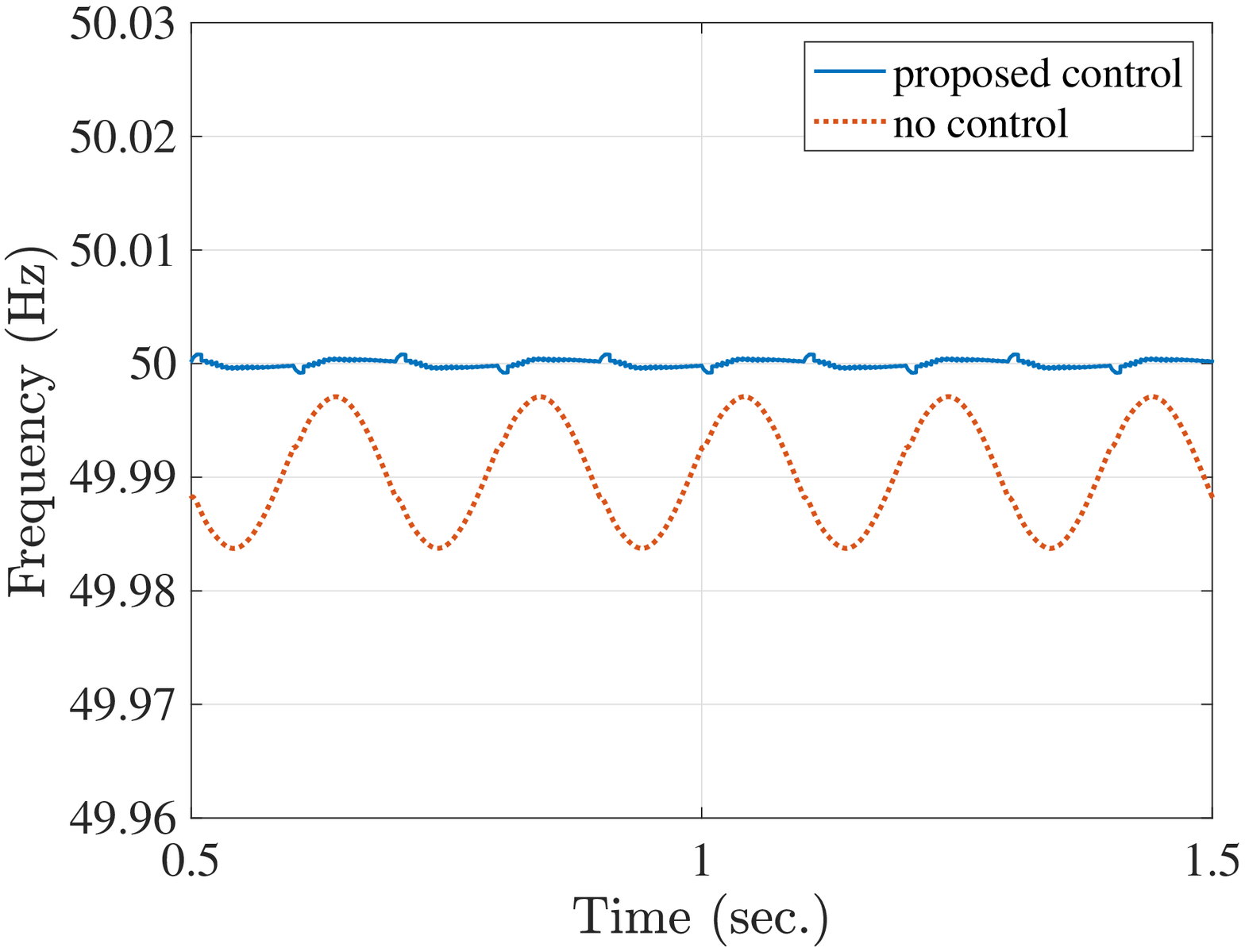}}
				\hfil
				\caption{Frequencies at (a) node 1, (b) node 2, and (c) node 3 with and without the proposed $\mu$AGC.}
				\label{fig: FastCtrPerformance}
			\end{figure*}

\subsubsection{Control Performance with Fast Renewable Fluctuations}
Suppose Load $1$ varies periodically and Figure \ref{fig: LoadFluctuate} shows the real power flowing to Load 1. The spikes in Figure \ref{fig: LoadFluctuate}(a) show  the load changing transients, and the zoomed-in version of one of the spikes is shown in Figure \ref{fig: LoadFluctuate}(b). As a result, the frequencies at the three nodes of Microgrid $1$ fluctuate periodically. This can be observed via the orange dashed curves in Figure \ref{fig: FastCtrPerformance}. 
The blue curves in Figure \ref{fig: FastCtrPerformance} show  the frequencies at the three nodes with fast load/renewable power fluctuations, when the proposed $\mu$AGC$_1$ is enabled. It can be seen that the magnitudes of the frequencies' fluctuations are significantly reduced by the proposed $\mu$AGC under load/renewable power fluctuations.

\begin{figure*}[hbt]
				\centering
				\subfloat[]{\includegraphics[width=0.25\linewidth]{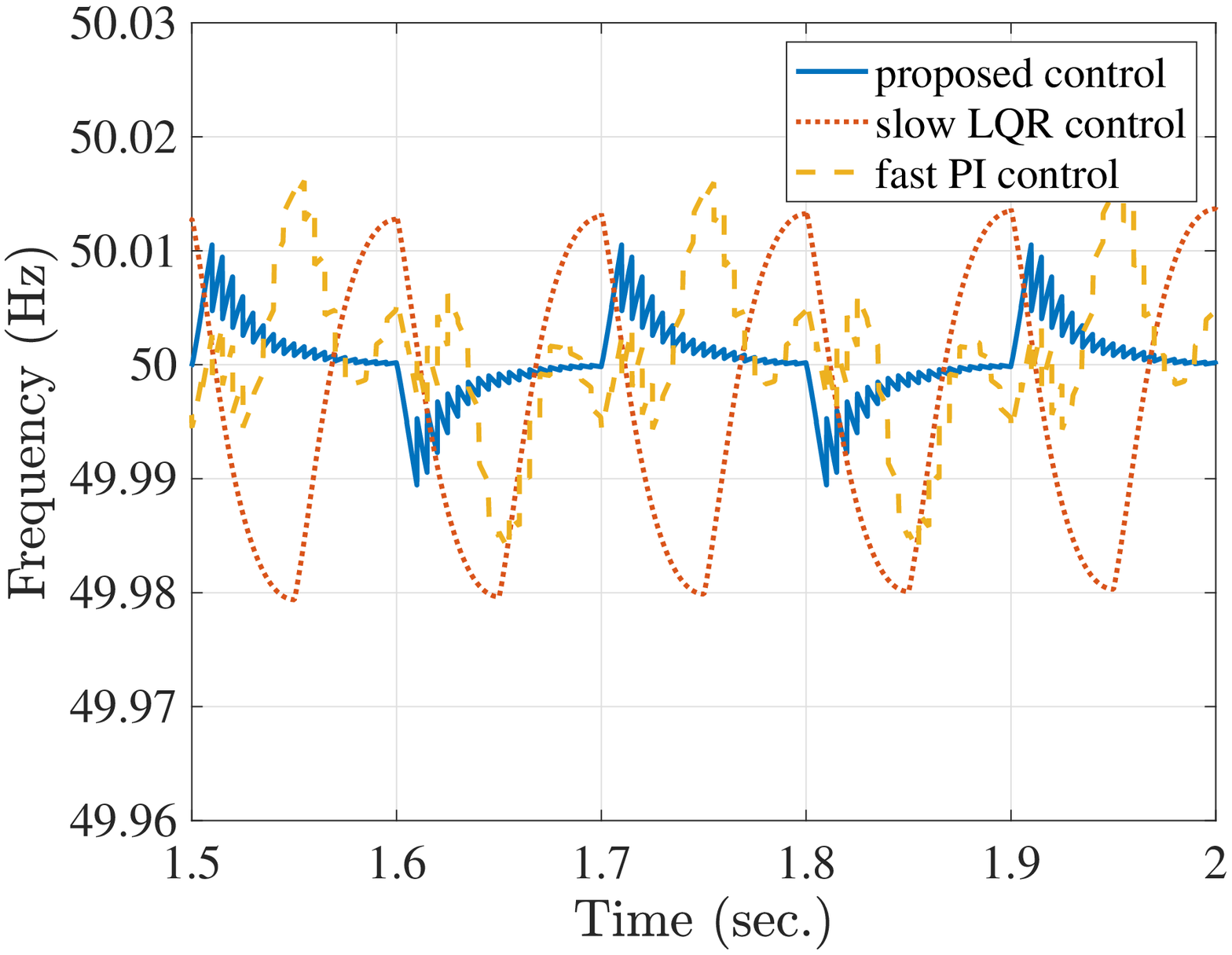}}
				\hfil
				\subfloat[]{\includegraphics[width=0.25\linewidth]{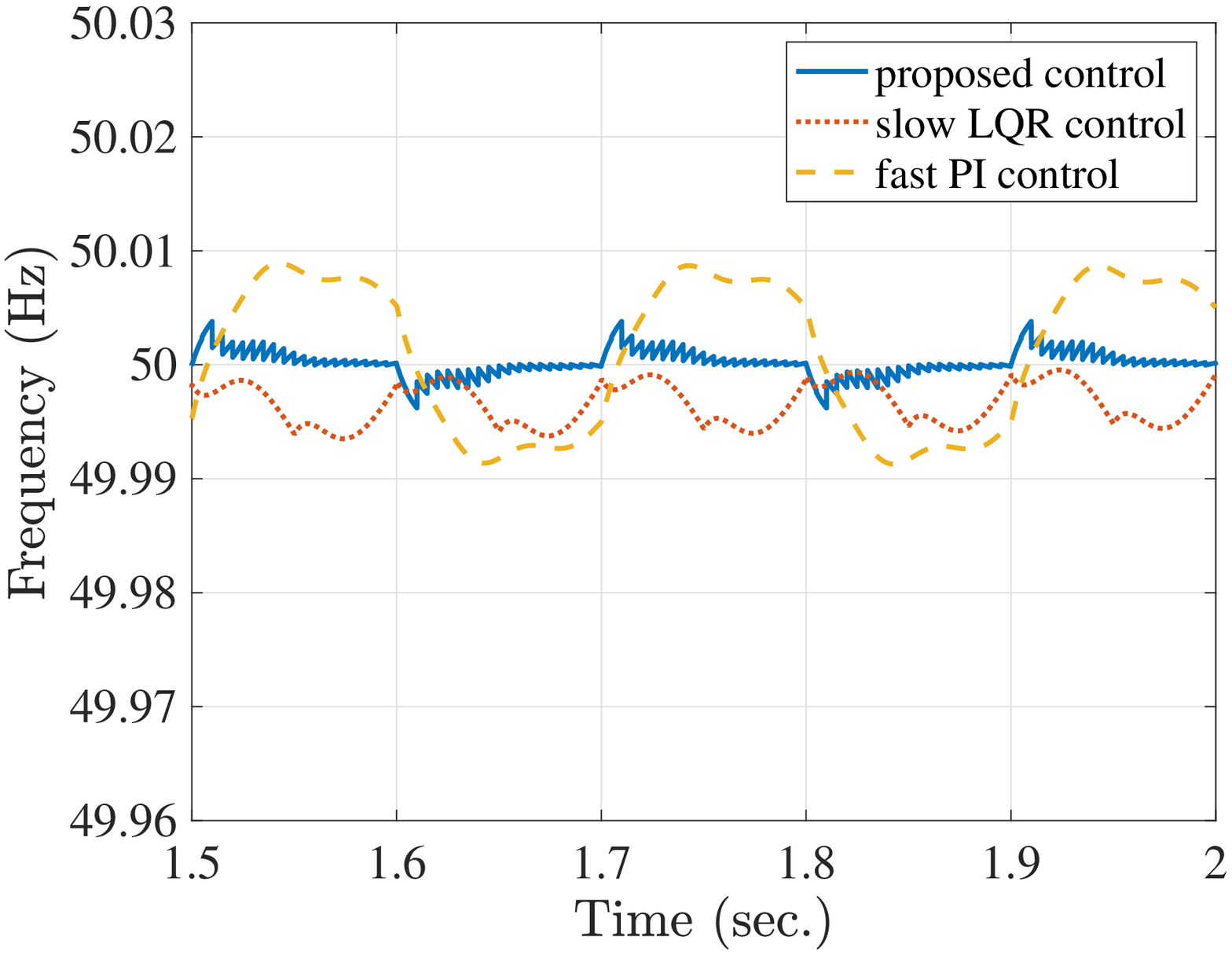}}
				\hfil
				\subfloat[]{\includegraphics[width=0.25\linewidth]{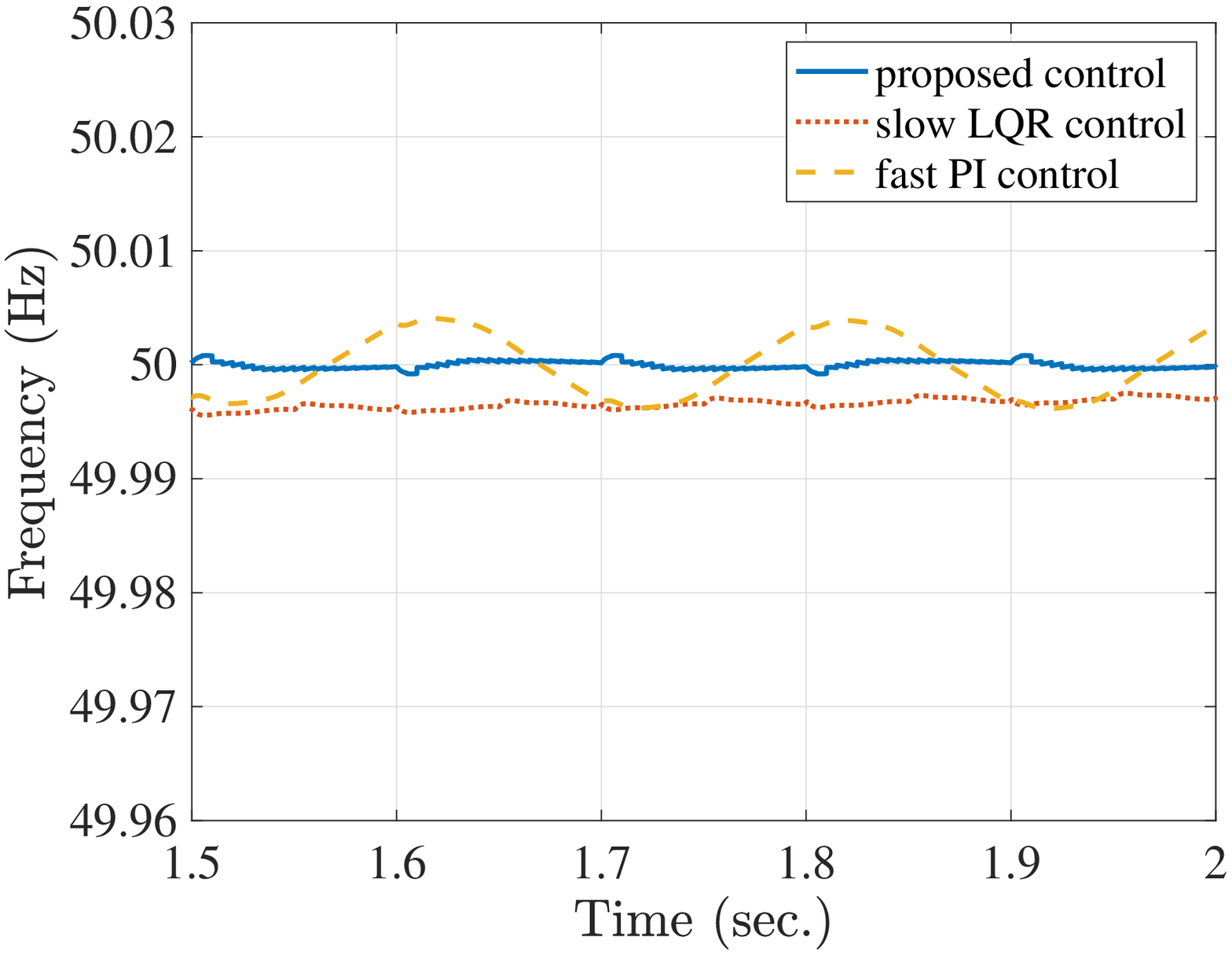}}
				\hfil
				\caption{Comparison of the proposed $\mu$AGC, the slow LQR-based controller, and the fast PI controller: frequencies at Nodes 1 (a), 2 (b), and 3 (c).}
				\label{fig: FastCtrPerformanceALL}
			\end{figure*}

\subsubsection{Comparison}
The proposed $\mu$AGC is compared with a slow LQR $\mu$AGC and a fast proportional–integral (PI) $\mu$AGC. The slow $\mu$AGC issues control commands every $0.1$ s, while the two fast $\mu$AGC issues control commands every $5$ ms. Figure \ref{fig: FastCtrPerformanceALL} summarizes the comparison results in response to the  load power fluctuations shown in Figure \ref{fig: LoadFluctuate}. In Figure \ref{fig: FastCtrPerformanceALL}, the orange dotted curves are the frequencies under the slow $\mu$AGC. It can be observed that the frequencies cannot be regulated. This is not surprising because the time interval between two successive control commends is the same as that between two successive disturbances. The yellow dashed curves are the frequency response with the fast PI $\mu$AGC. Since the fast frequency changes cannot be measured accurately, the frequencies under the fast PI controller still cannot be regulated. As the proposed $\mu$AGC does not rely on the availability of fast frequency measurements, the frequencies can be regulated tightly around $50$ Hz under the fast load/renewable power fluctuations, as it is shown by the blue curves in Figure \ref{fig: FastCtrPerformanceALL}.

\subsection{Cyber Attack Detection and Corrective Control}
The proposed scheme for cyber resilient frequency regulator is tested under both renewable/load power fluctuations and cyber attacks.

\begin{figure}[h!]
				\centering
				\subfloat[]{\includegraphics[width=0.45\linewidth]{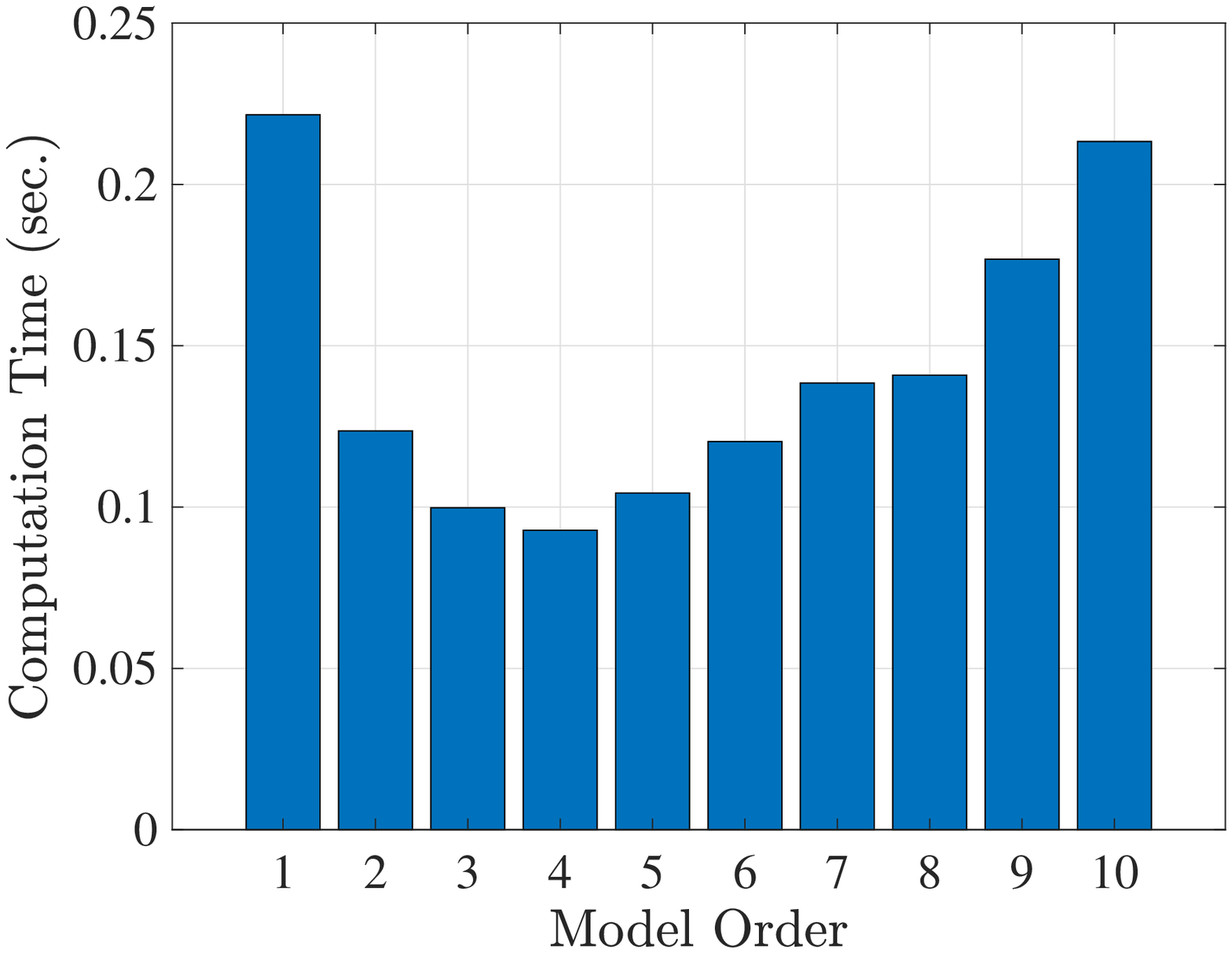}}
				\hfil
				\subfloat[]{\includegraphics[width=0.45\linewidth]{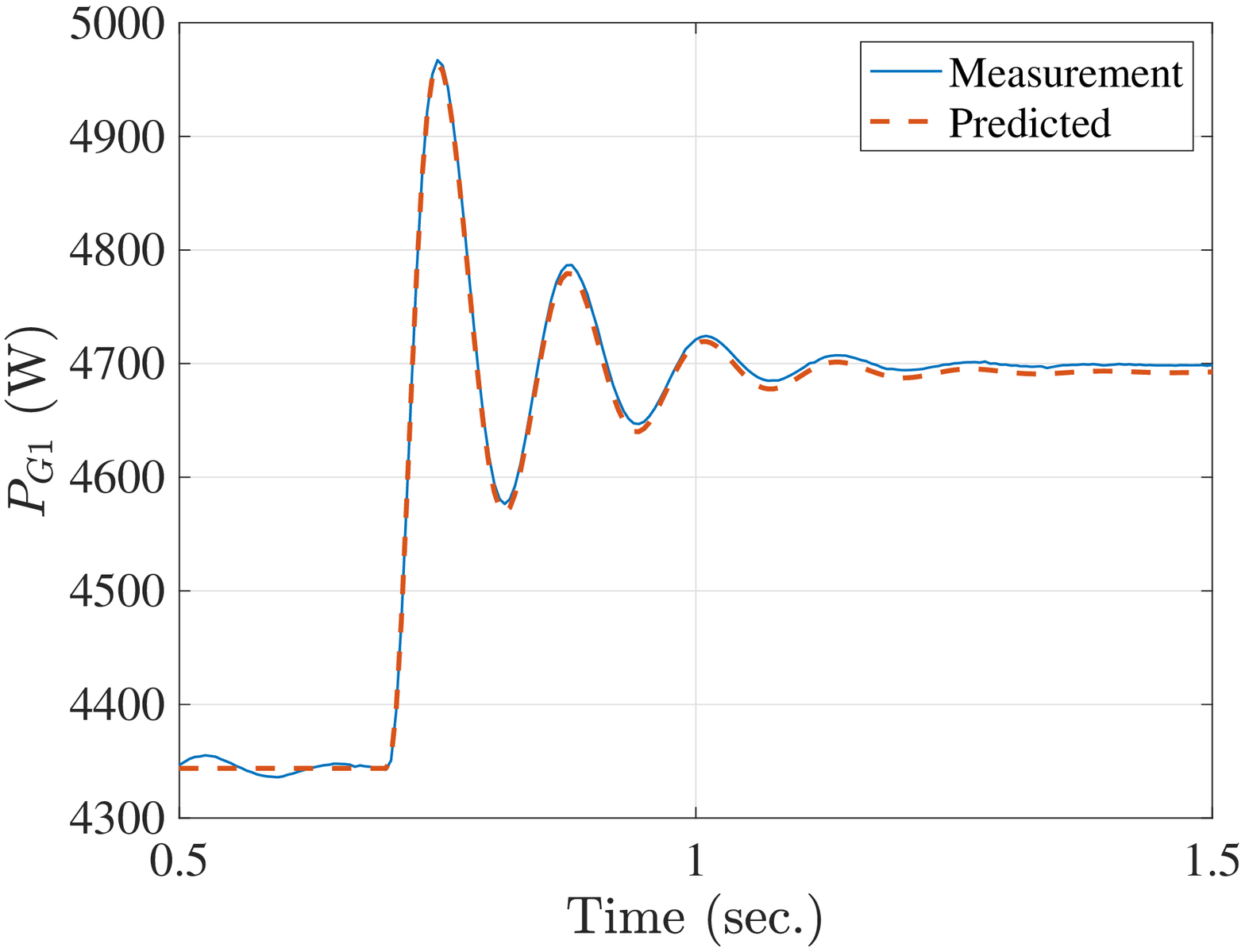}}
				\hfil
				\caption{(a) Computation time of the data-driven approach with different model orders; and (b) Actual measurements and predicted measurements}
				 \label{fig:SystemID}
			\end{figure}

\subsubsection{Cyber Attack Detection}
\label{ssub:attack_detection}
One prerequisite for implementing Algorithm \ref{alg:1} is that the matrices $A'$, $B'$, and $C'$ are available. In this paper, these matrices are learned from the measurements of $\Delta \boldsymbol{\omega}_{\text{s}}$ and $\Delta \mathbf{P}_{\text{G}}$ based on the system identification procedure. It takes $1.43$ seconds to decide the optimal order $d^*$ and to learn the discrete model offline using a MacBook Pro with a 2.6 GHz Dual-Core Intel i5 Processor. The sampling rate of the discrete model is $0.005$ seconds. The computation time of learning the models with different orders are shown in Figure \ref{fig:SystemID}(a). Next, we show the performance of the model learned by changing the setpoint of one IBR. Without the load/renewable power fluctuations, the set point $\Delta \omega_{\text{s}1}$ 
has a step change at time $t=0.7$ s. The blue curve in Figure \ref{fig:SystemID}(b) represents the evolution of the measurements on $P_{\text{G}1}$ under the step change. With the model learned, the evolution of $P_{\text{G}1}$ can be predicted almost instantaneously. The orange dashed curve in Figure \ref{fig:SystemID}(b) represents the prediction. These two curves in Figure \ref{fig:SystemID}(b) almost overlap with each other, suggesting that the state-space model learned can accurately predict the response of the real power measurements $P_{\text{G}1}$.


\begin{figure}[h!]
				\centering
				\subfloat[]{\includegraphics[width=0.45\linewidth]{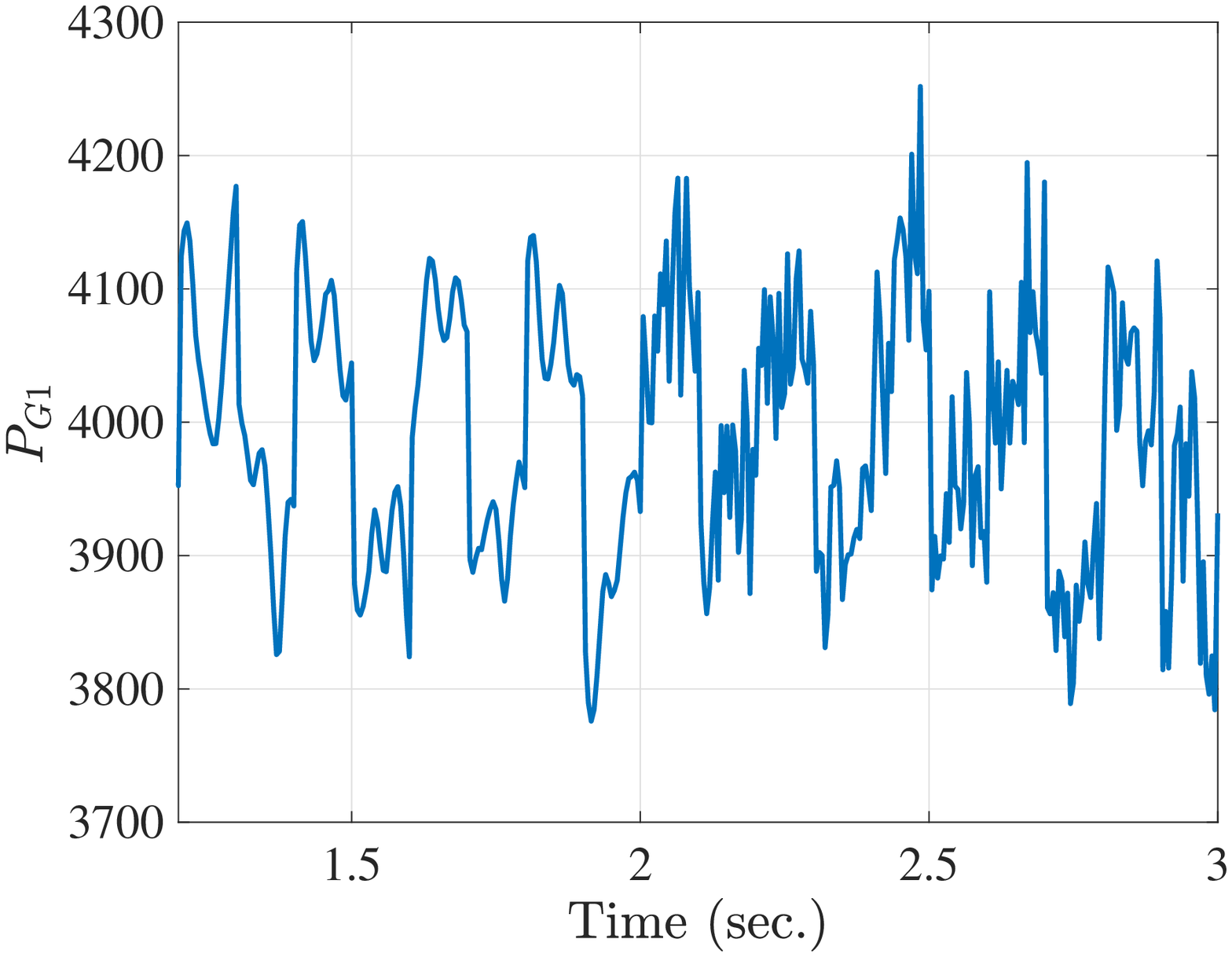}}
				\hfil
				\subfloat[]{\includegraphics[width=0.45\linewidth]{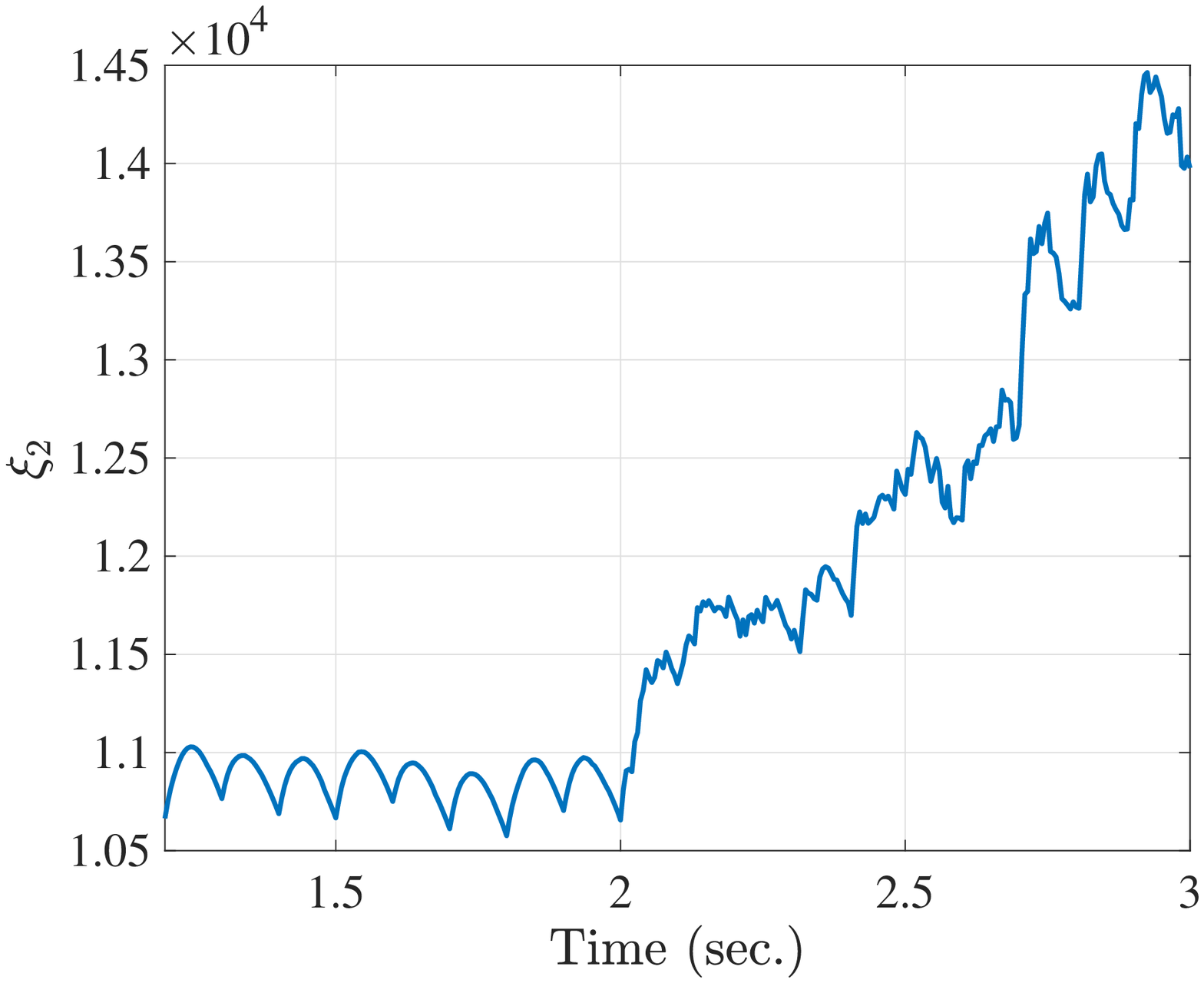}}
				\hfil
				\caption{Detection of the noise injection attack: (a) the real power measurement $P_{\text{G}i}$ in the time domain; and (b) the evolution of cyber attack indicator $\xi_2$.}
				\label{fig: NoiseInjection}
			\end{figure}

\begin{figure}[h!]
				\centering
				\subfloat[]{\includegraphics[width=0.45\linewidth]{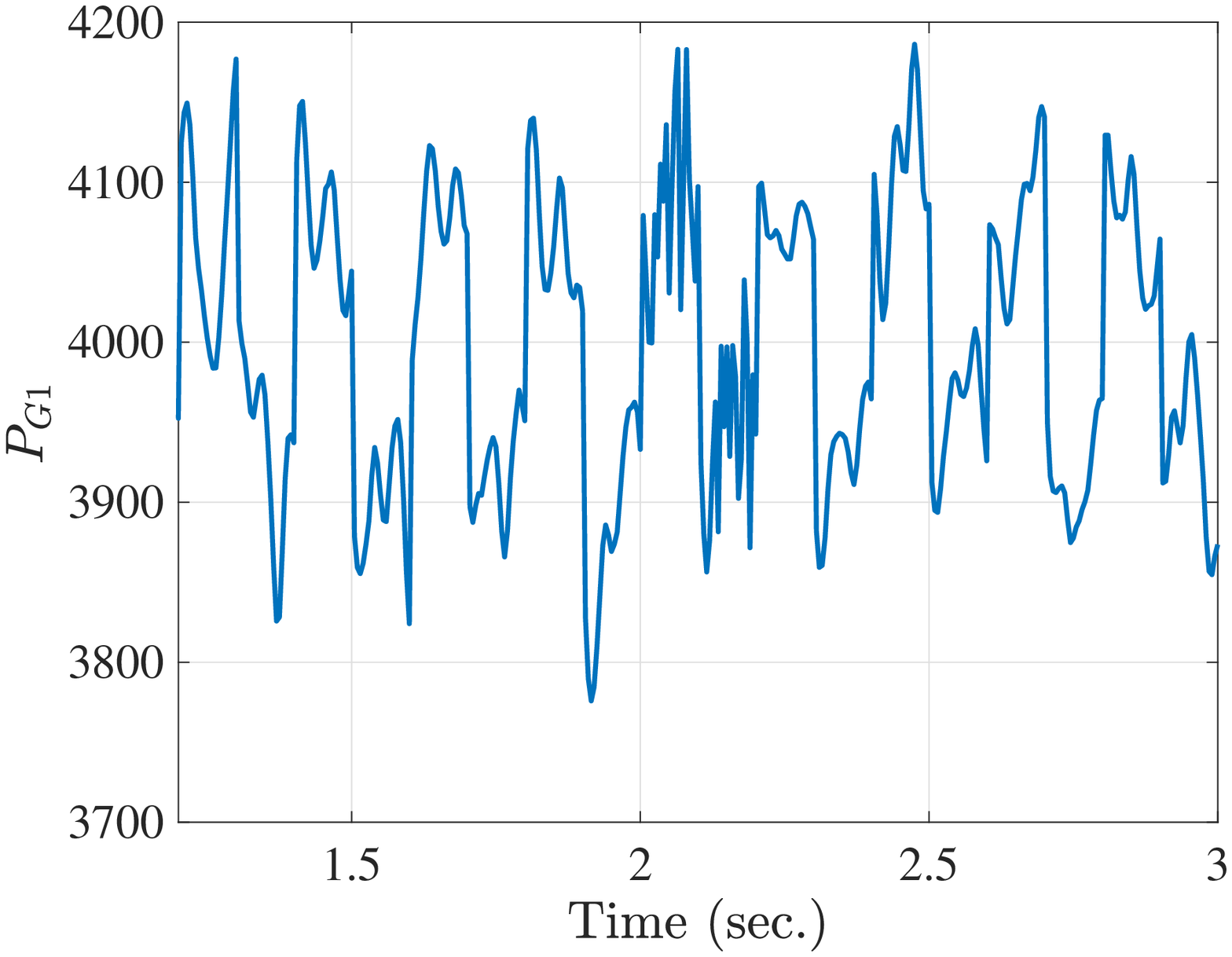}}
				\hfil
				\subfloat[]{\includegraphics[width=0.45\linewidth]{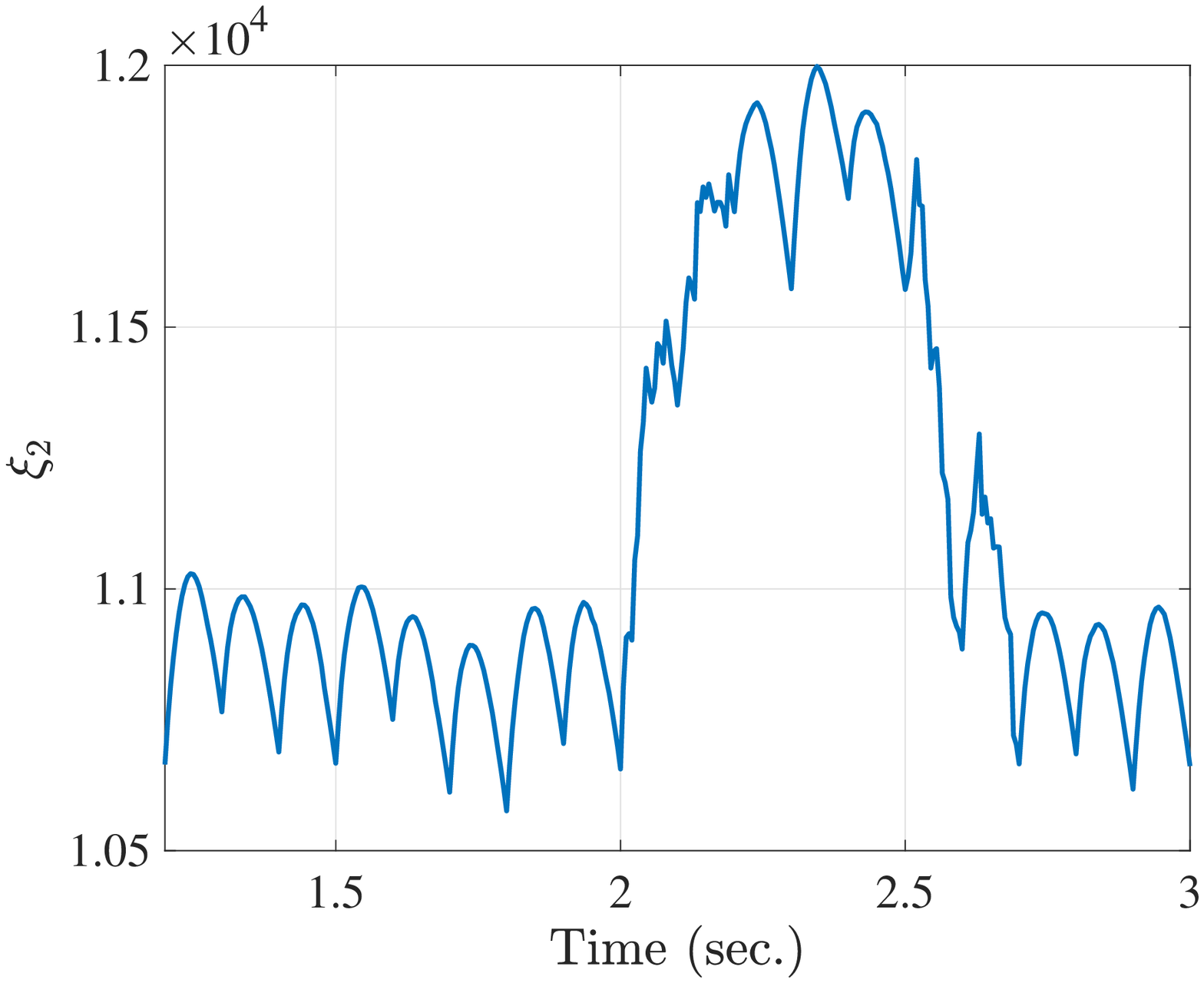}}
				\hfil
				\caption{Detection of the temporary noise injection attack: (a) the real power measurement $P_{\text{G}i}$ in the time domain; and (b) the evolution of cyber attack indicator $\xi_2$.}
				\label{fig: NoiseInjection_temp}
			\end{figure}

\begin{figure}[h!]
				\centering
				\subfloat[]{\includegraphics[width=0.45\linewidth]{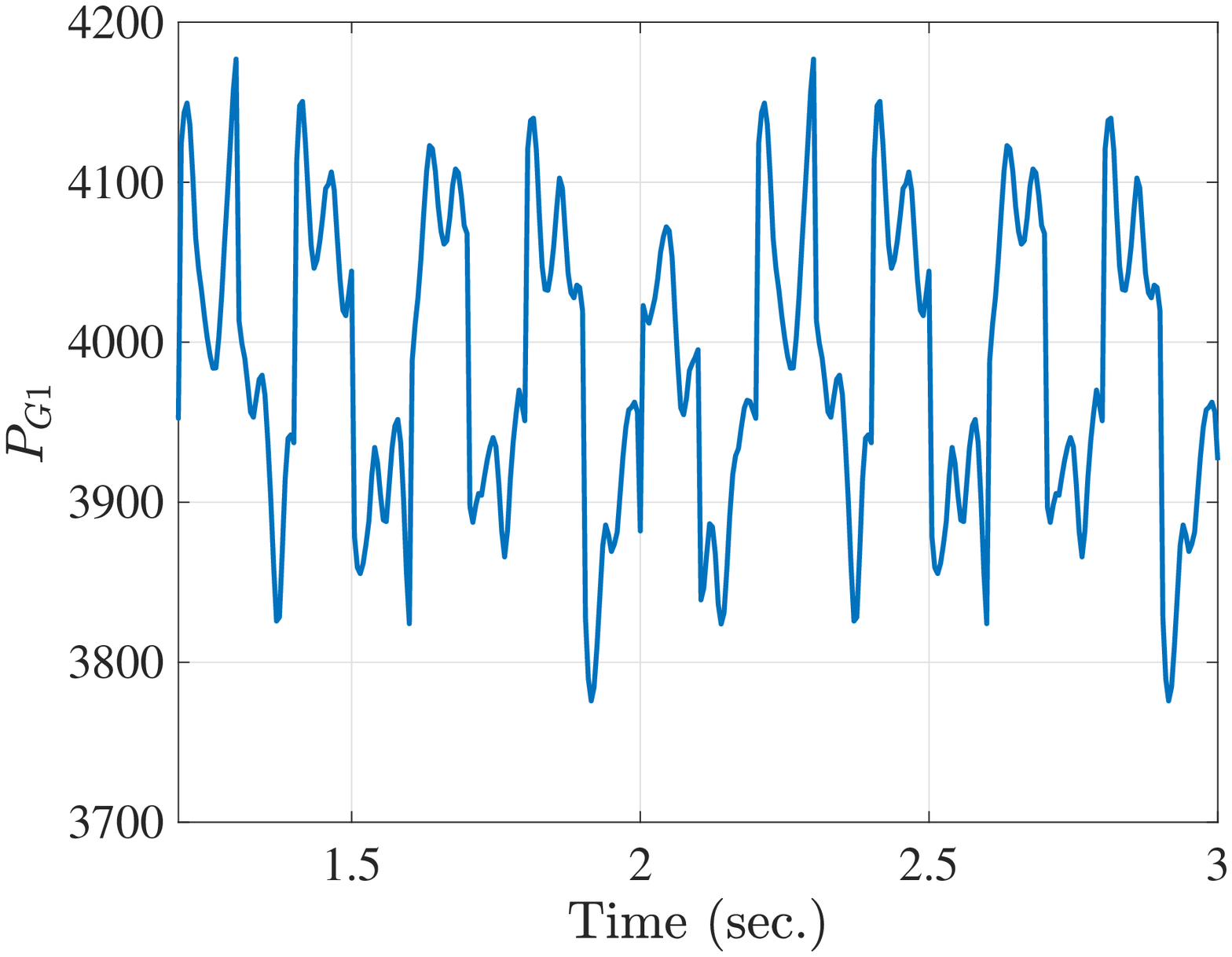}}
				\hfil
				\subfloat[]{\includegraphics[width=0.4\linewidth]{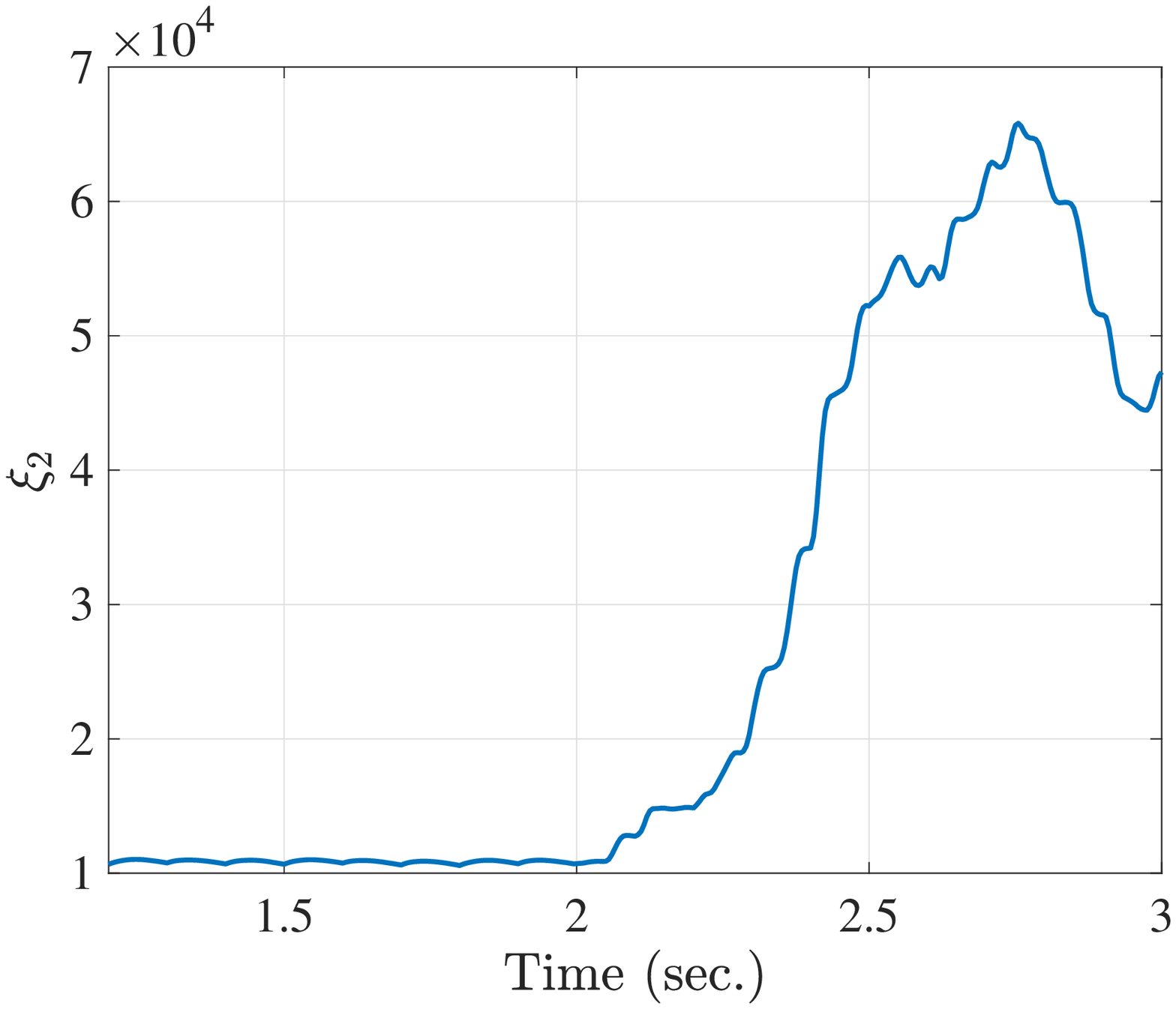}}
				\hfil
				\caption{Detection of the replay attack: (a) the real power measurement $P_{\text{G}i}$ in the time domain; and (b) the evolution of cyber attack indicator $\xi_2$.}
				\label{fig: Replay}
			\end{figure}

The performance of Algorithm \ref{alg:1} is tested under the noise injection attack and the replay attack. In the simulation, $W$ in Algorithm \ref{alg:1} is set to $100$. Figures \ref{fig: NoiseInjection} and \ref{fig: NoiseInjection_temp} show the performance of Algorithm \ref{alg:1} under the noise injection attack. In Figure \ref{fig: NoiseInjection}(a), the noise injection attack on $P_{\text{G}1}$ is launched at time $t=2$ s. Such an attack causes the attack indicator $\xi_2$ to shoot up and triggers the cyber attack alarm. In Figure \ref{fig: NoiseInjection_temp}(a), a temporary noise injection attack starts at $t=2$ s and ends at $t=2.2$ s. As shown in Figure \ref{fig: NoiseInjection_temp}(b), it takes around $0.025$ seconds for the indicator to exceed its normal peak value, after the attack occurs. After the attack disappears, it takes around $0.485$ seconds for the indicator to fall below its normal peak value.
Figure \ref{fig: Replay} illustrates the performance of of Algorithm \ref{alg:1} under the replay attack. A replay attack on the $P_{\text{G}1}$ occurs at time $t=2$ s. Figure \ref{fig: Replay}(a) shows the $P_{\text{G}1}$ measurements under such an attack. From Figure \ref{fig: Replay}(a), it is hard to determine when the cyber anomaly occurs by simply eyeballing the $P_{\text{G}1}$ measurement curve. However, Figure \ref{fig: Replay}(b) which shows the evolution of the attack indicator $\xi_2$ explicitly suggests that the attack occurs around $t=2$ s.

\subsubsection{Observer-based Cyber Attack Correction}
\label{ssub:observer}
\begin{figure}
    \centering
    \includegraphics[width =0.5\linewidth]{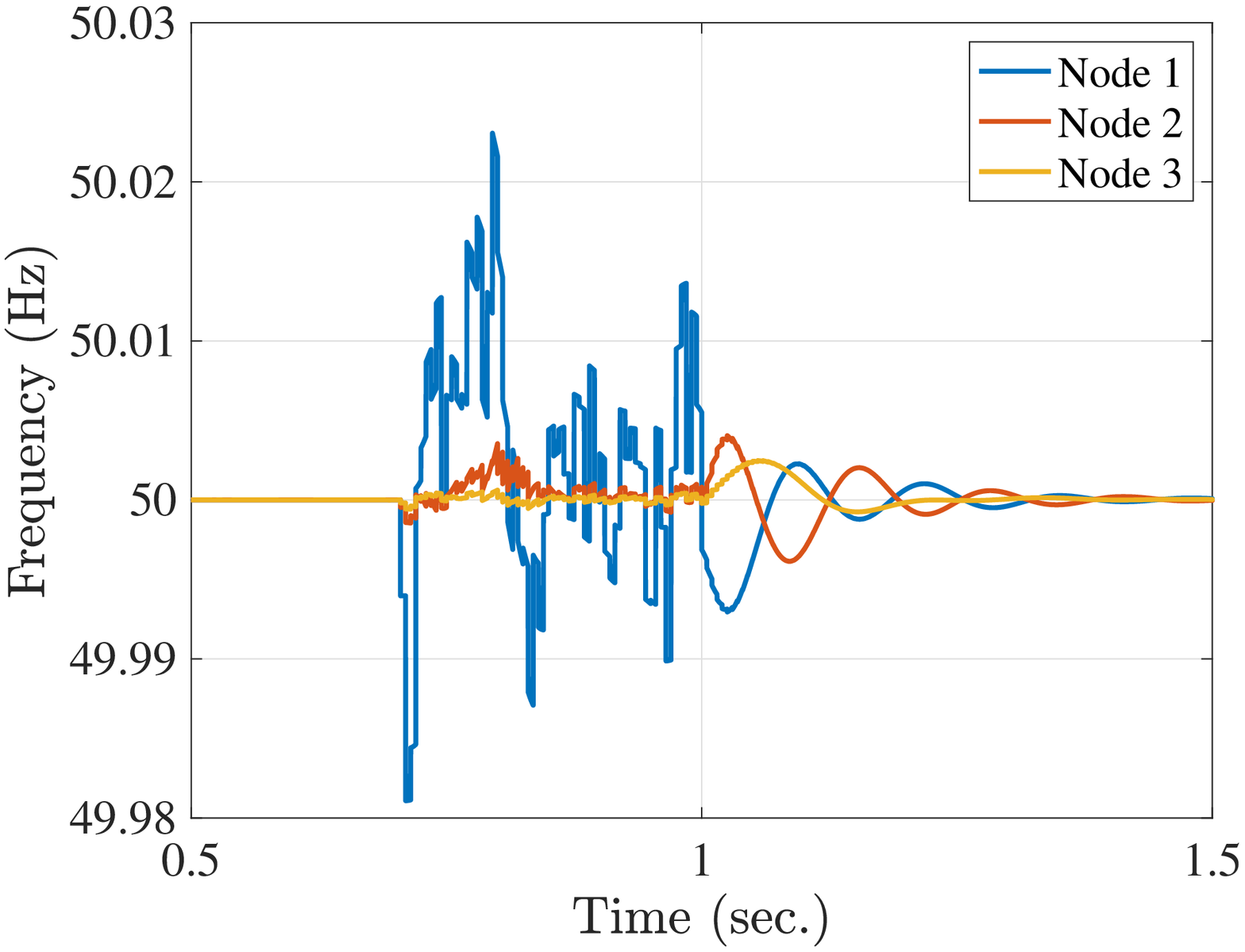}
    \caption{Performance of the observer-based controller}
    \label{fig:observer_ctr}
\end{figure}

Figure \ref{fig:observer_ctr} demonstrates the performance of the observer-based corrective controller. The noise injection attacks occur at the sensor measuring $P_{\text{G}1}$ at Microgrid 1 at $t=0.7$ s. Consequently, frequencies at the three nodes of Microgrid 1 become noisy. At $t= 1$ s, the observer-based corrective controller is enabled, and the frequencies are stabilized.

\subsubsection{Collaborative Cyber Attack Correction}
\label{ssub:network}
\begin{figure}[h!]
				\centering
				\subfloat[]{\includegraphics[width=0.45\linewidth]{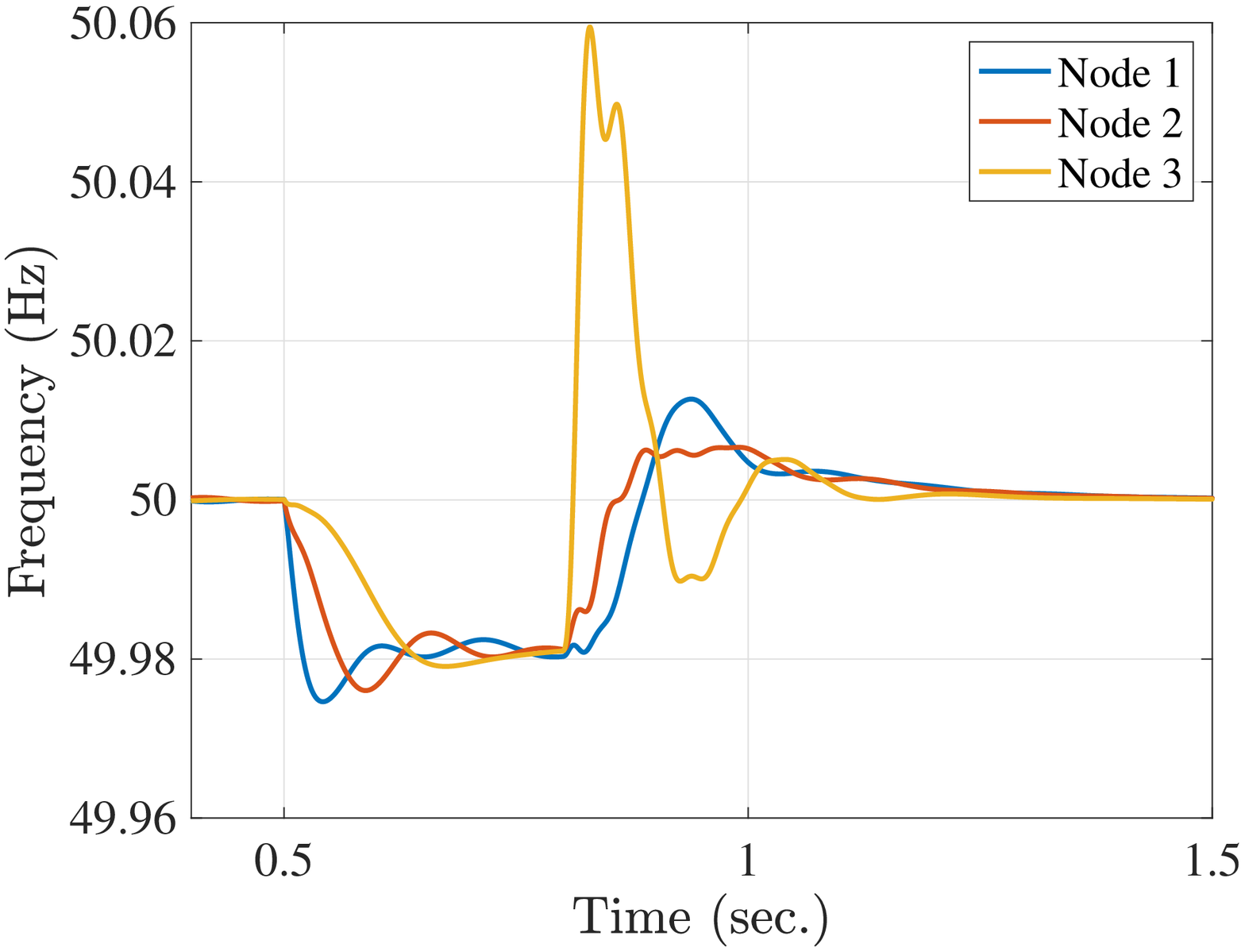}}
				\hfil
				\subfloat[]{\includegraphics[width=0.45\linewidth]{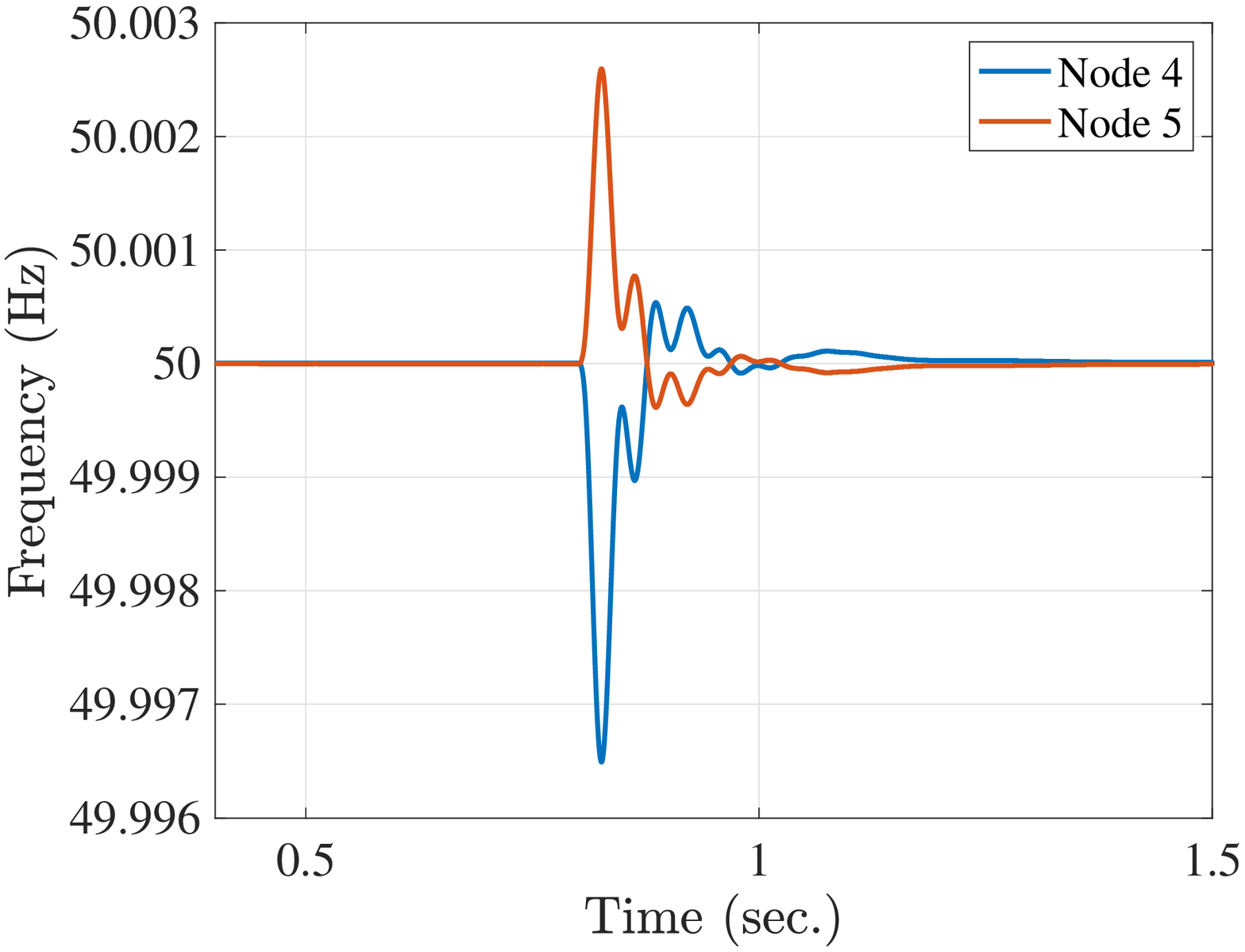}}
				\hfil
				\caption{Performance of collaborative corrective control scheme: (a) frequencies at microgrid 1 under cyber attack; and (b) frequencies at microgrid 2.}
				\label{fig: CollaborativeCtr}
			\end{figure}

The observer-based corrective control is blind to renewable/load power fluctuations. If Microgrid 2 is available, once cyber attacks at Microgrid 1 are detected, Microgrid 1 can disable $\mu$AGC$_1$ and turn on the switches of the tie line that connects Microgrid 2. The $\mu$AGC$_2$ can regulate the frequencies of both Microgrids 1 and 2. Figure \ref{fig: CollaborativeCtr} illustrates such a process. Suppose that at $t=0.5$ s, $\mu$AGC$_1$ is disabled due to a cyber attack detected. Load $1$ also increases at $t=0.5$ s, leading frequencies at Microgrid 1 to decrease. To recover the frequencies to $50$ Hz, Microgrid 1 networks with Microgrid 2 via the tie line at $t=0.7$ sec. It can be seen from Figure \ref{fig: CollaborativeCtr} that the three frequencies at Microgrid 1 are brought back to $50$ Hz, and the event of networking Microgrids 1 and 2 only causes negligible frequency ripples at Microgrid 2.

\begin{figure*}[bt]
				\centering
				\subfloat[]{\includegraphics[width=0.25\linewidth]{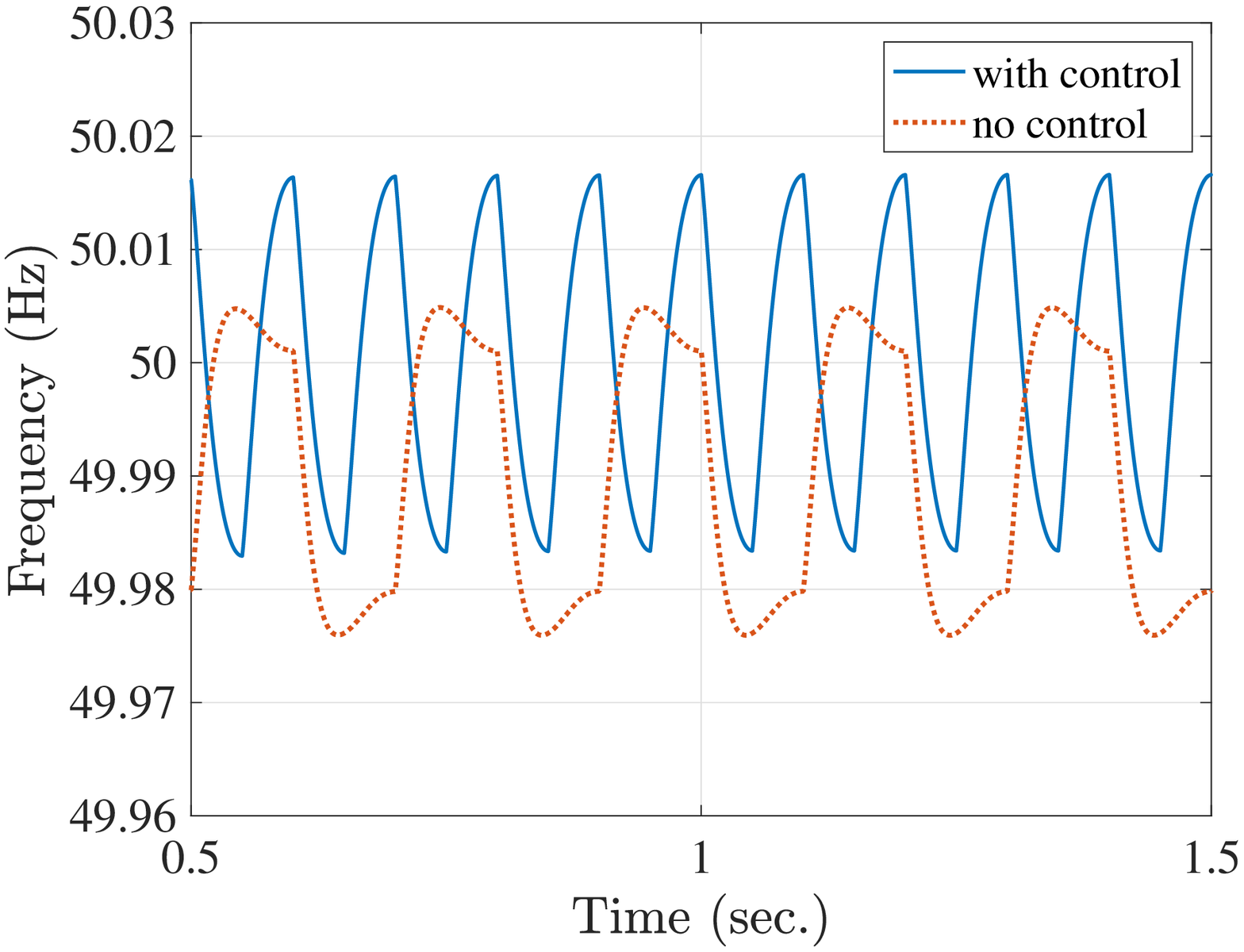}}
				\hfil
				\subfloat[]{\includegraphics[width=0.25\linewidth]{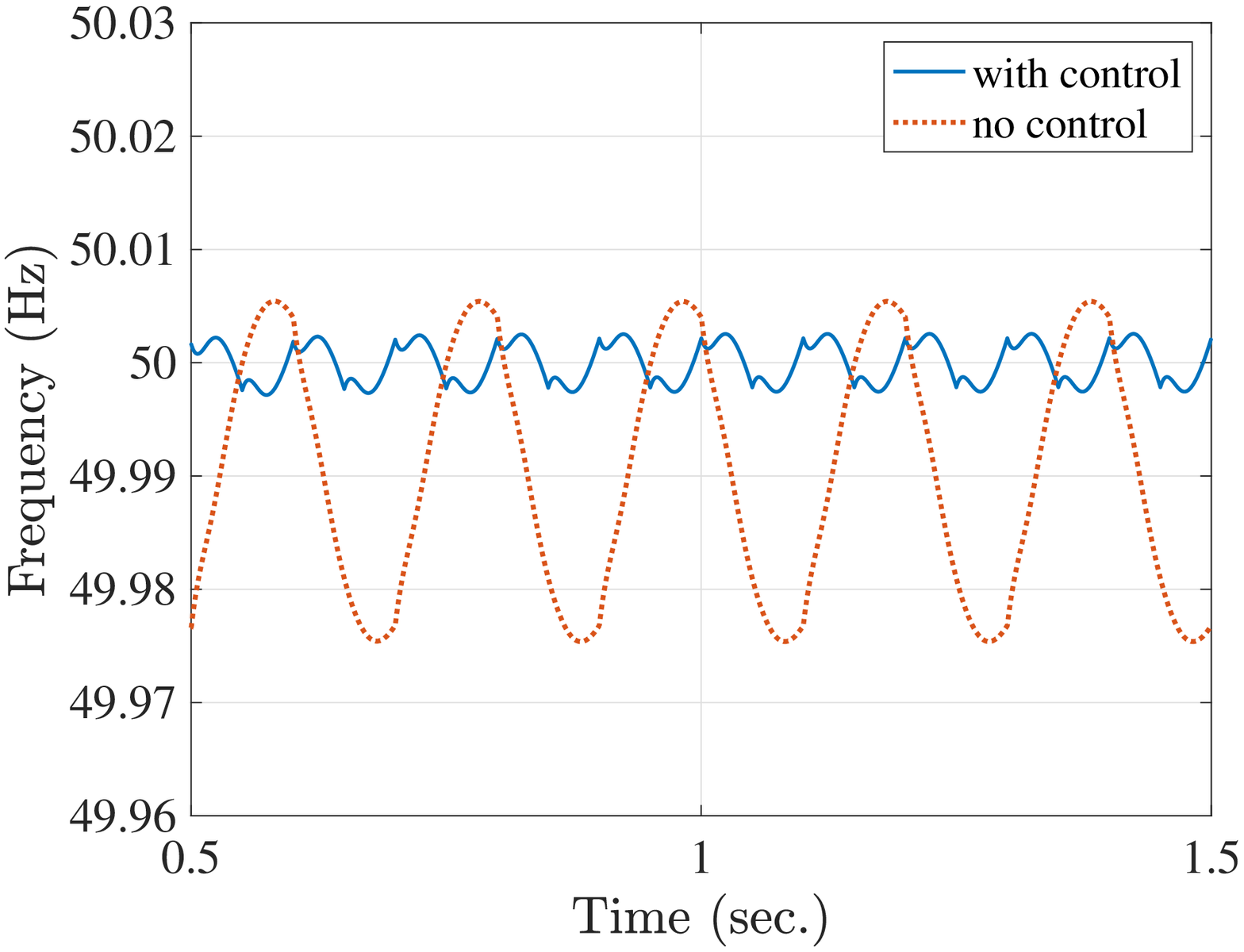}}
				\hfil
				\subfloat[]{\includegraphics[width=0.25\linewidth]{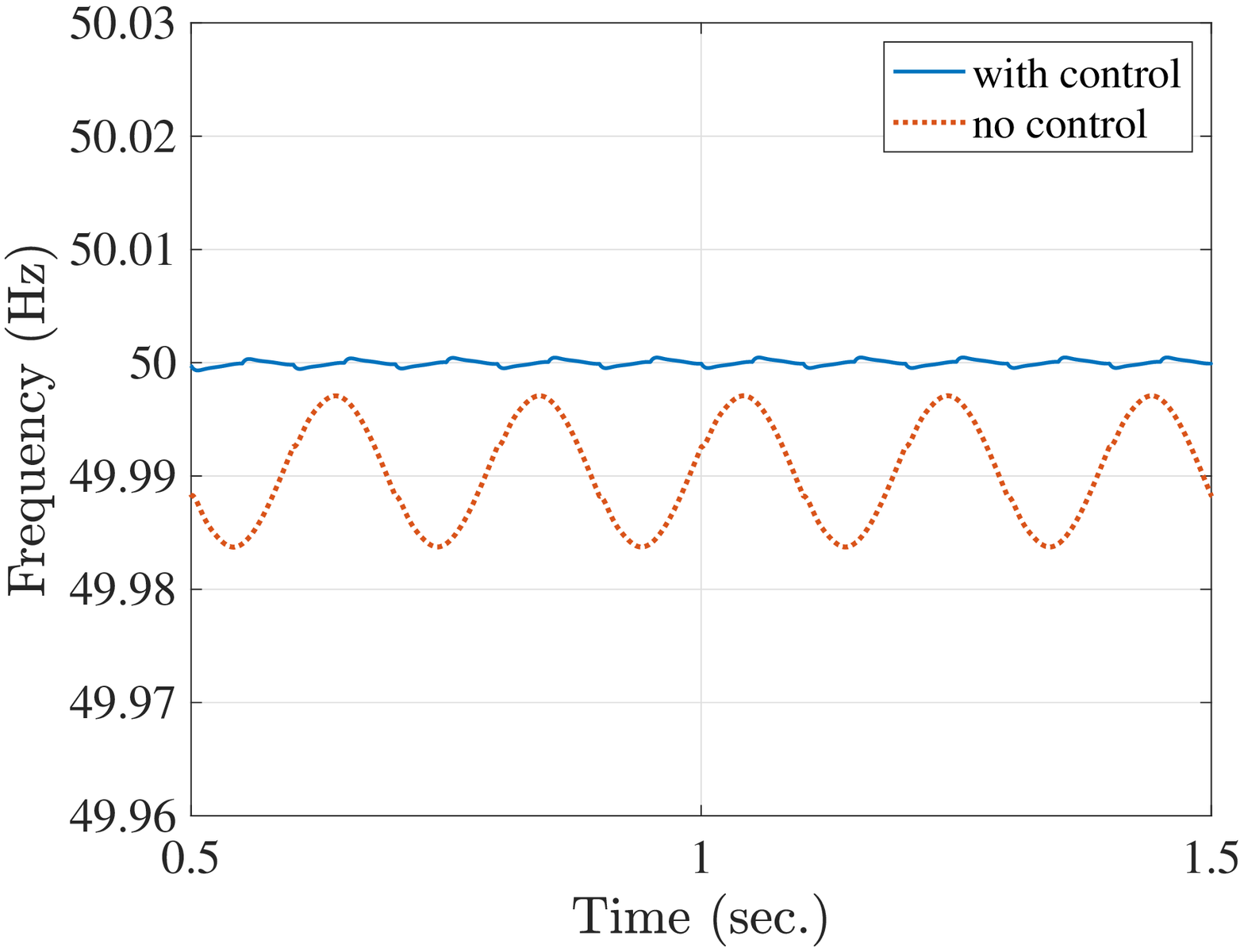}}
				\hfil
				\caption{\textcolor{black}{Performance of collaborative control in terms of regulating frequencies at IBR 1 (a), IBR 2 (b), and IBR 3 (c): the solid blue curves are frequency trajectories with the collaborative control, and the dashed orange curves are frequency trajectories without the collaborative control.}}
    \label{fig: f123_fast}
			\end{figure*}

\begin{figure}
    \centering
    \includegraphics[width = 3in]{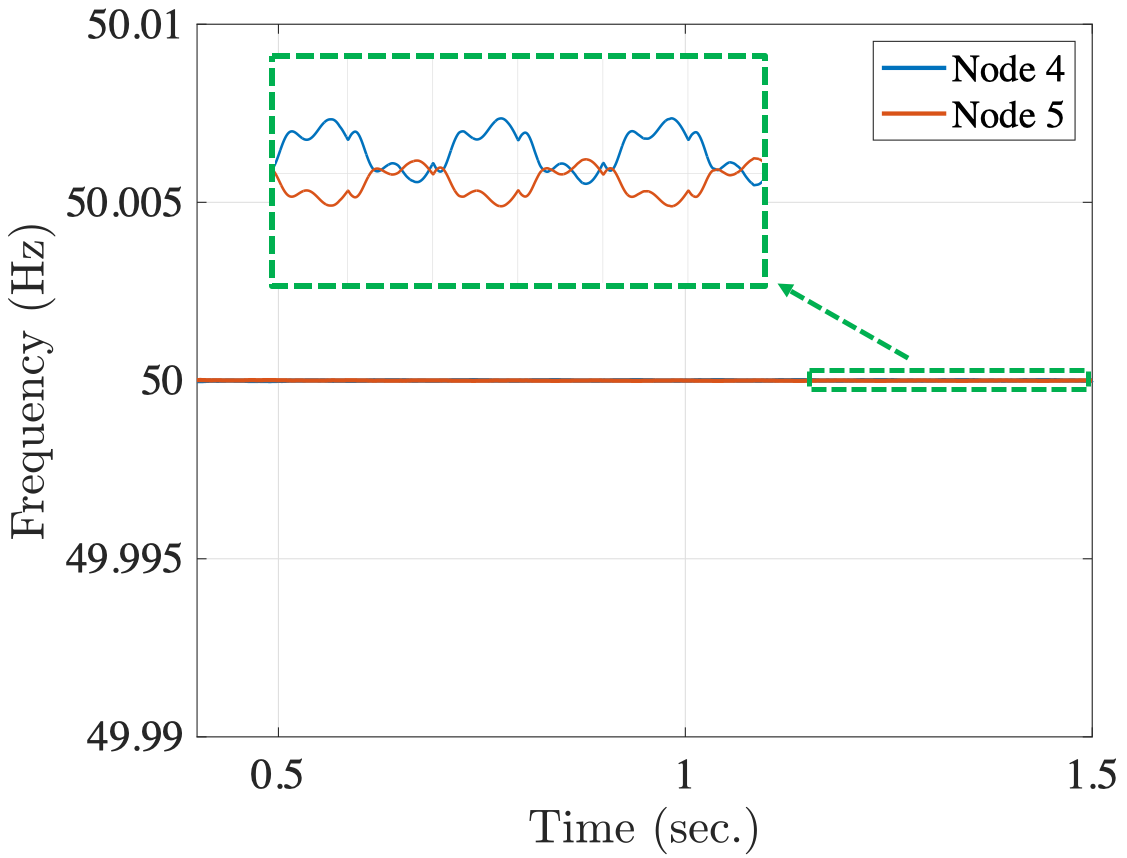}
    \caption{\textcolor{black}{Frequencies at Microgrid 2}}
    \label{fig:MG2_freq_fast}
\end{figure}

\textcolor{black}{$\mu$AGC$_2$ only serves as a temporary frequency regulator for Microgrid 1 when Microgrid 1 is under cyber attacks. Suppose that $\mu$AGC$_1$ for Microgrid 1 is disabled, and $\mu$AGC$_2$ for Microgrid 2 is enabled. Load at node 1 fluctuates fast according to Figure \ref{fig: LoadFluctuate}.
Figure \ref{fig: f123_fast} compares frequencies with and without the collaborative control. It can be seen that while $\mu$AGC$_2$ does not play a big role of regulating the frequency at node 1, it indeed improves the frequency profiles at nodes 2 and 3 by limiting the frequencies of nodes 2 and 3 within a narrower range. Figure \ref{fig:MG2_freq_fast} shows that $\mu$AGC$_2$ can regulate the frequencies at IBRs 4 and 5 tightly around $50$ Hz in the presence of fast load fluctuations at node 1. Since the control performance of enabling $\mu$AGC$_2$ alone is not as good as that of enabling both $\mu$AGCs, both $\mu$AGCs should be enabled if no cyber attacks are detected.}

\begin{figure}
    \centering
    \includegraphics[width = 2.5in]{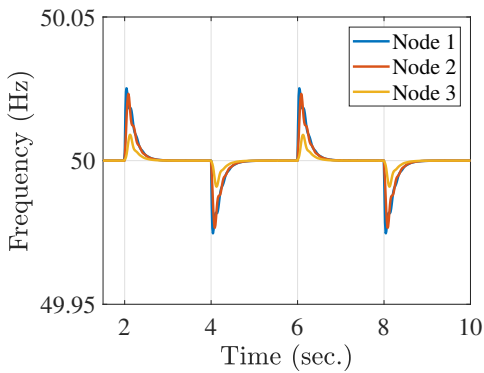}
    \caption{\textcolor{black}{Frequencies at Microgrid 1 under slow load fluctuations}}
    \label{fig:MG2_freq_slow}
\end{figure}

\textcolor{black}{It is worth noting that with $\mu$AGC$_1$ disabled, $\mu$AGC$_2$ can still regulate frequencies in Microgrid $1$ under slow load fluctuations. Figure \ref{fig:MG2_freq_slow} shows the frequencies at IBRs 1, 2, and 3 with the load at node 1 that changes every $2$ s, when $\mu$AGC$_1$ is disabled. It can be observed that the frequencies at IBRs $1$, $2$, and $3$ are tightly regulated around their nominal values by $\mu$AGC$_2$ under the slow load fluctuations at node 1.}

\begin{figure}[bt]
				\centering
				\subfloat[]{\includegraphics[width=0.5\linewidth]{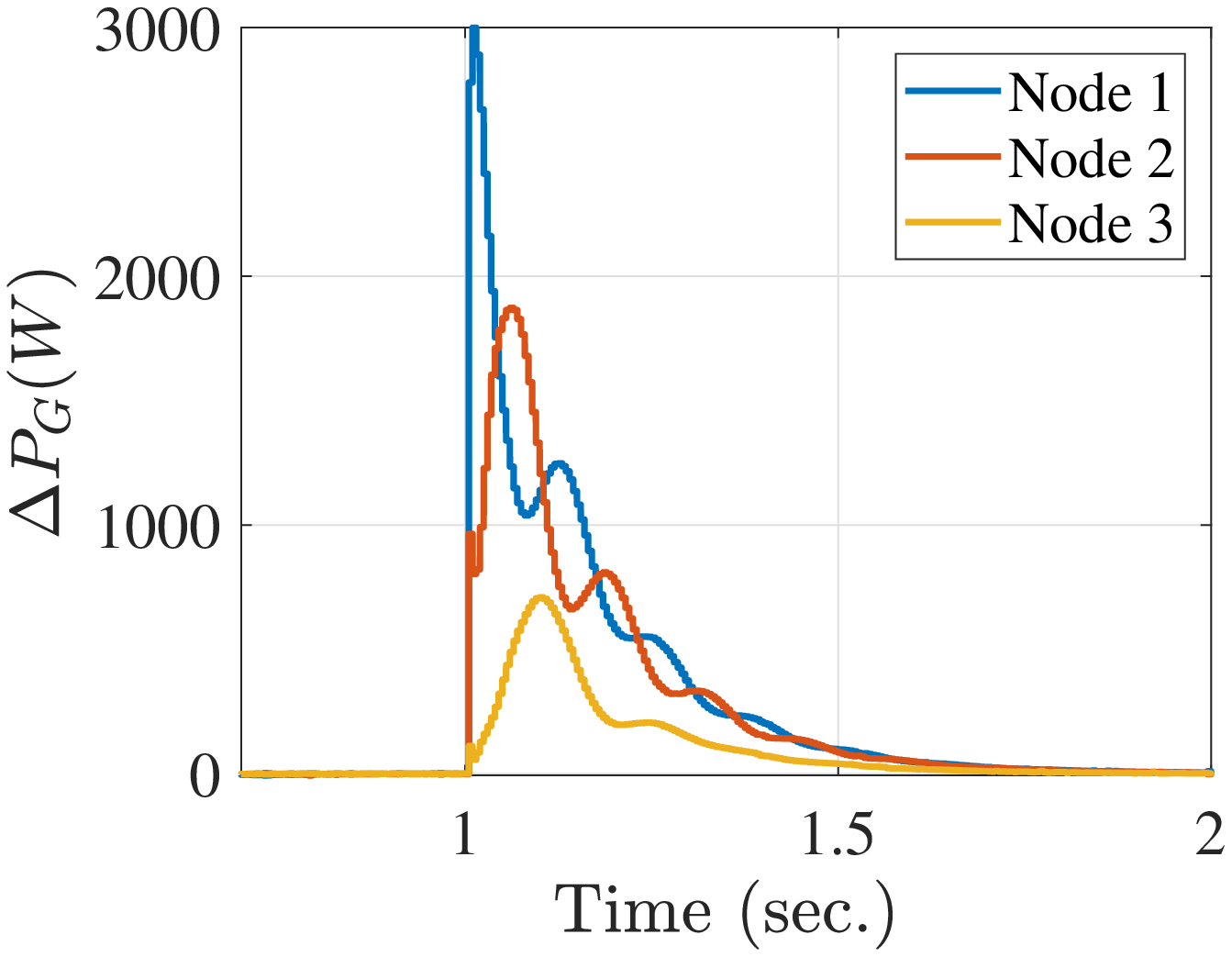}}
				\hfil
				\subfloat[]{\includegraphics[width=0.5\linewidth]{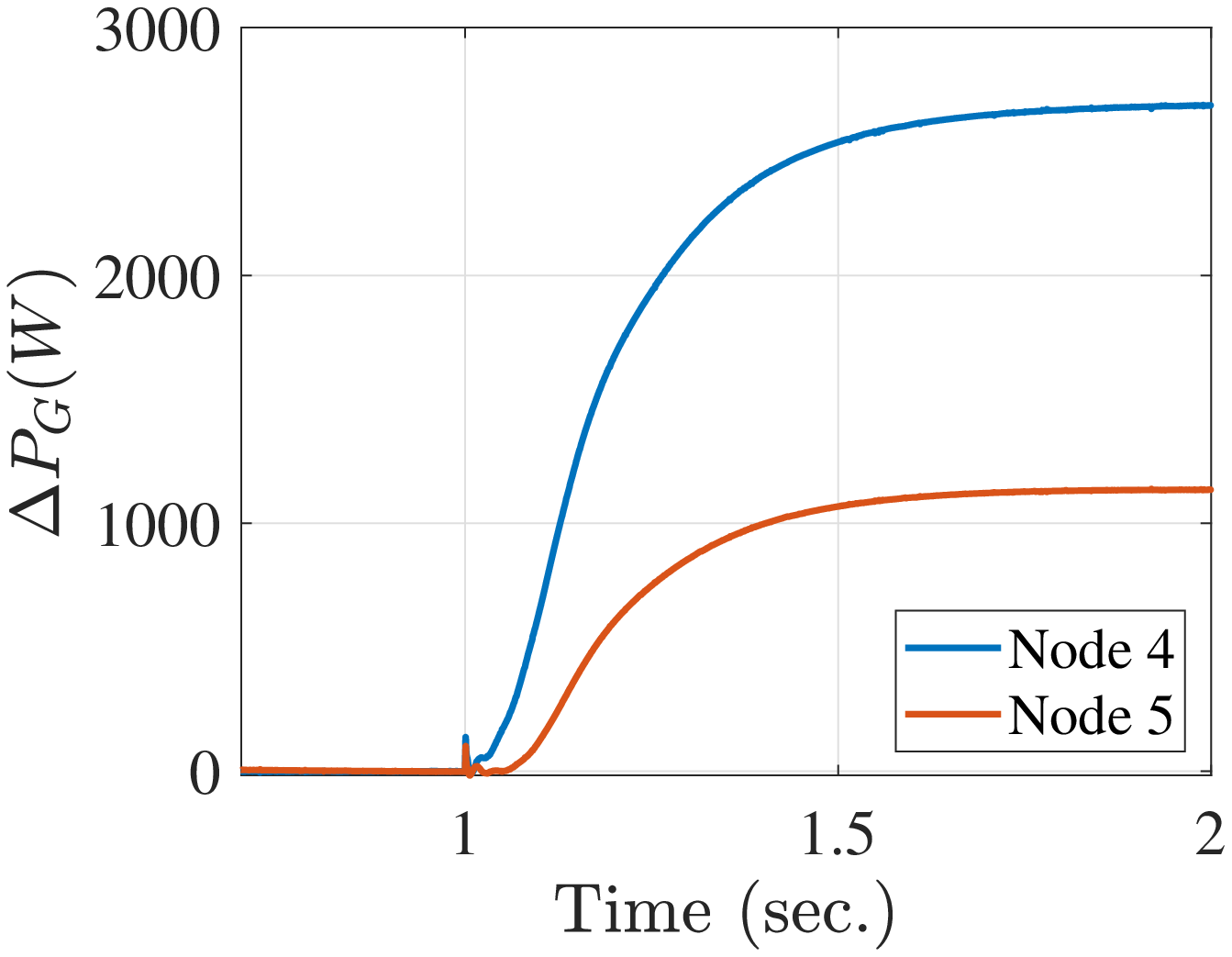}}
				\hfil
				\caption{\textcolor{black}{(a) $\Delta P_{\text{G}i}$ for $i=1,2,3$; (b) $\Delta P_{\text{G}i}$ for $i=4,5$}}
    \label{fig: justification}
			\end{figure}

\textcolor{black}{The performance of the collaborative control can be justified by examining $\Delta P_{\text{G}i}$ in \eqref{eq:z_dynamics}.
When the load at node 1 increases at $1$ s and the two microgrids are networked, Figure \ref{fig: justification} visualizes $\Delta P_{\text{G}i}$ for $i = 1, 2, \ldots, 5$. In Figure \ref{fig: justification}-(a), it can be observed that the steady-state values of $\Delta P_{\text{G}i}$ for $i= 1,2,3$ return zero, which means that no extra real power is produced by IBRs that do not participate in $\mu$AGC$_2$, i.e., IBRs 1, 2 and 3, in order to balance the increased amount of load. The increased amount of load is balanced by the IBRs that participate in $\mu$AGC$_2$, i.e., IBRs 4 and 5. Since IBRs 1, 2 and 3 do not participate in $\mu$AGC$_2$, $\Delta \omega_{\text{s}i}=0$ for $i=1,2,3$. In addition, Figure \ref{fig: justification}-(a) shows $\Delta P_{\text{G}i}=0$ for $i=1,2,3$ in steady state. Based on \eqref{eq:z_dynamics}, $\dot{z_i}$ equals zero for $i= 1,2,3$. Because of Observation 1, $\Delta \omega_i$ is zero in steady state for $i=1,2,3$. This explains why the frequencies in Microgrid 1 is regulated by $\mu$AGC$_2$, while $\mu$AGC$_2$ does not take any measurements from Microgrid 1. Note that such performance of the collaborative control is built upon the assumption that no congestion occurs in the tie line between Microgrids 1 and 2. Suppose that the congestion occurs, the increase amount of real power cannot be balanced by the IBRs in Microgrid 2. In such a case, $\mu$AGC$_2$ cannot regulate frequencies in Microgrid 1 to the nominal value.}

\section{Conclusion}
In this paper, a cyber-resilient $\mu$AGC is proposed. A new microgrid modeling approach is introduced. The modeling approach exploits the rank-deficient property of IBR dynamics. Such modeling approach leads to a fast $\mu$AGC design that requires no frequency measurements that may not be available in practice under fast load/renewable power fluctuations. An end-to-end cyber security solution to FDI attack detection and mitigation is developed for the proposed $\mu$AGC. The proposed cyber-resilient $\mu$AGC is tested in a system of two microgrids. Simulation shows the proposed $\mu$AGC can regulate microgrid frequencies under fast renewable/load power fluctuations, and its cyber security solution can detect and mitigate FDI attacks. Future work will investigate the impact of microgrid nonlinearity on the cyber-resilient $\mu$AGC design and test and validate the proposed $\mu$AGC in a large-scale system of AC microgrids.

\bibliographystyle{IEEEtran}
\bibliography{ref.bib}
\appendices
\end{document}